\documentclass[aps,prx,twocolumn,showpacs,superscriptaddress,longbibliography]{revtex4-1}

\usepackage{graphicx}
\usepackage{amsmath}
\usepackage{amssymb}
\usepackage[normalem]{ulem}
\usepackage{xcolor}

\usepackage[english]{babel}
\usepackage[colorlinks]{hyperref}

\begin{document}

\title{Multiple abrupt phase transitions in urban transport congestion}

\author{Aniello Lampo}
        \affiliation{Internet Interdisciplinary Institute (IN3), Universitat Oberta de Catalunya, Barcelona, Catalonia, Spain}
\author{Javier Borge-Holthoefer}
        \affiliation{Internet Interdisciplinary Institute (IN3), Universitat Oberta de Catalunya, Barcelona, Catalonia, Spain}
\author{Sergio G\'{o}mez}
        \affiliation{Departament d'Enginyeria Inform\`{a}tica i Matem\`{a}tiques, Universitat Rovira i Virgili, Tarragona, Catalonia, Spain}
\author{Albert Sol\'{e}-Ribalta}\email{asolerib@uoc.edu}
        \affiliation{Internet Interdisciplinary Institute (IN3), Universitat Oberta de Catalunya, Barcelona, Catalonia, Spain}
        \affiliation{URPP Social Networks, University of Zurich, Zurich, Switzerland}

\begin{abstract}
During the last decades, the study of cities has been transformed by new approaches combining engineering and complexity sciences. Network theory is playing a central role, facilitating the quantitative analysis of crucial urban dynamics, such as mobility, city growth or urban planning. In this work, we focus on the spatial aspects of congestion. Analyzing a large amount of real city networks, we show that the location of the onset of congestion changes according to the considered urban area, defining, in turn, a set of congestion regimes separated by abrupt transitions. To help unveiling this spatial dependencies of congestion (in terms of network betweenness analysis), we introduce a family of planar road network models composed by a dense urban center connected to an arboreal periphery. These models, coined as GT and DT-MST models, allow us to analytically, numerically and experimentally describe how and why congestion emerges in particular geographical areas of monocentric cities and, subsequently, to describe the congestion regimes and the factors that promote the appearance of their abrupt transitions. We show that the fundamental ingredient behind the observed abrupt transitions is the spatial separation between the urban center and the periphery, and the number of separated areas that form the periphery. Elaborating on the implications of our results, we show that they may have influence in the design and optimization of road networks regarding urban growth and the management of daily traffic dynamics.
\end{abstract}

\maketitle

\section{Introduction}

Cities evolved, and continue evolving, into different organizational patterns as a consequence of historical, political or financial circumstances and continuous optimization \cite{Bettencourt2010,Batty2012}. City constituents usually interact with one another, requiring a complex systems perspective to describe the observed phenomena. Urban transportation networks, such as metro, bus lines or road networks, are paradigmatic examples of fields that have attracted the interest of the physics community for, at least, two decades \cite{Helbing1998, Helbing2001, Guimera2002}.

For the topic of interest here, the analysis of road networks, different approaches have been taken depending on their individual characteristics and its purpose. Dynamics on inter-urban roads (a.k.a.\ arterial roads, or high capacity roads), characterized by long segments and limited interconnection, is governed by the interaction between cars circulating within. Thus, most of the phenomena arising in this context (congestion, traffic waves, phantom traffic jams, slower-is-faster effect, etc.), can be explained by considering vehicles moving on independent roads, e.g., using car-following models and fluid dynamics \cite{helbing2015traffic, Kerner2004, Treiber2013, Chowdhury2000, Nagatani2002}, or the fundamental diagram of traffic flow which relates density and flux of vehicles \cite{godfrey1969mechanism, daganzo2008analytical}. On the contrary, in intra-urban dynamics, vehicles continuously relocate to different road segments during their trajectory, which makes that most of the phenomenology observed is governed by the road network structure modelling the dependencies among these segments. In this situation, complex networks theory plays a salient role. Theoretical models developed along these lines, such as the classical UTA model \cite{Kerner2004}, have helped to understand several important characteristics of such systems, e.g., urban mobility \cite{Barthelemy2009,Balcan2009}, traffic \cite{Guimera2002,Echenique2005,ling2010global,sole2018decongestion,akbarzadeh2018communicability,verbavatz2019critical,gao2019effective,chen2020traffic}, road usage \cite{Wang2012,Strano2012}, and network collapse \cite{olmos2018macroscopic,zhang2019scale,zeng2019switch}.

These two types of road networks (high capacity and urban roads), which are usually analyzed independently, are increasingly entangled as cities sprawl over suburban areas \cite{Strano2012}. As shown in similar systems, when different structural patterns are combined to compose a new system, we may observe emergent phenomena induced by this entanglement such as double phase transitions \cite{Colomer2014,Mata2015,Hackett2016}, special cases of the Braes paradox \cite{sole2016congestion}, or increase in network resilience \cite{de2014navigability}. This evidences that current approaches are too narrow to fully understand road network structure and dynamics. Up to date, only few works have studied the urban transportation networks from this intertwined perspective. In \cite{Kirkley2018} the authors describe the structural properties of systems where arterial and local urban roads share the same geographic space. This model allows them to explain the universal shape of the experimentally observed betweenness distribution of the road networks in a large set of worldwide cities.

In this work, we focus on the spatial aspects of congestion on monocentric cities where arterial roads and urban local ones operate on separate geographic spaces. Specifically, we assume that local roads are basically located at the city center, and arterial roads at the urban periphery. Although this may seem an oversimplification of the real situation, and probably neither of both extreme cases (complete overlap or complete separation) are fully compatible with observations, as we will see, it will suffice to unveil several interesting properties of road networks related to the congestion phenomena.

With the previous considerations in mind, we develop a family of planar road network models with equivalent structural properties, a dense urban center and a periphery with arboreal structure. We start by analyzing the most idealized situation of the model family that we call the Grid-Tree (GT) model. This model is able to reproduce previous results in terms of betweenness distribution \cite{Kirkley2018}, and at the same time offers a considerable advantage in terms of analytical tractability. The GT-model equations evidence that cities may experience a set of multiple abrupt phase transitions in the spatial localization of congested areas. These, in turn, define a set of congestion regimes: emergence of congestion in the city center, in its periphery, or in urban arterial roads. We further elaborate on these transitions by conducting analysis on a more general model family, the DT-MST model (from Delaunay Triangulation and Maximum Spanning Tree). Surprisingly, abrupt transitions hold for such a general context as well, allowing us to conclude that they follow from the way different road types, located on separated spatial areas, are entangled to form an integrated road transportation system.

After the effort to characterize the transitions on synthetic networks, we turn our attention to the analysis of real road networks. For that empirical analysis, we first present an automatic and unsupervised method to characterize such regimes and transitions on almost a hundred cities. Results show that the predicted abrupt transitions exist in real cities and that our model is practically able to predict the exact number of them.

The manuscript is organized as follows. In Sec.~\ref{sec:BeetwDist}, we discuss the importance of the betweenness centrality in urban settings, its characteristics when the network is embedded in Euclidean spaces, and we relate our work to recent literature. In Sec.~\ref{sec:GTmodel}, we develop our GT-model and analyze it in terms of state-of-the-art results related to the betweenness distribution and its spatial behavior. Section~\ref{sec:AnBeetwCalc} is devoted to the derivation, and subsequent validation, of the analytical expressions for the betweenness of several distinguished nodes of the GT-model. In Sec.~\ref{sec:CongestionRegimes} we derive, using the previous analytical calculations, the abrupt transitions which define the congestion regimes. In Sec.~\ref{sec:RandomPlanarModels}, we introduce the general DT-MST model and discuss how its properties are related to the origin of the transitions. The existence of these different congestion regimes is validated with empirical road networks associated to real cities in Sec.~\ref{sec:RealCities}. Finally, in Sec.~\ref{sec:Discussion} we discuss on the relevance of the transitions in real cities and the applicability of our results, and Sec.~\ref{sec:Conclusions} contains some concluding remarks.

\section{Betweenness distribution in cities}
\label{sec:BeetwDist}

Betweenness, initially introduced in the social sciences \cite{Freeman1977,Freeman1978}, is a centrality measure of network constituents (nodes and edges) which quantifies their importance, in terms of the amount of paths crossing them. Besides sociology \cite{goh2003betweenness,everett2005ego,leydesdorff2007betweenness,kourtellis2013identifying,de2014role,de2014role}, betweenness centrality has been used in many other problems of interest: community detection \cite{girvan2002community,newman2004finding}, epidemic spreading \cite{matamalas2018effective}, percolation \cite{xia2010cascading,wandelt2018comparative}, and targeted attacks to networks \cite{holme2002attack,wang2009cascade,iyer2013attack,bellingeri2014efficiency}, to name a few \cite{barthelemy2004betweenness,sole2014centrality,wang2015ability,de2015ranking}. Yet probably, one of the most prominent applications is in the analysis of traffic and routing \cite{Guimera2002,Echenique2005,zhao2005onset,yan2006efficient,liu2007method,dolev2010routing,dong2012enhancing,tan2014traffic,sole2016congestion,sole2016model,sole2018decongestion,manfredi2018mobility,sole2019effect}.

Betweenness centrality is implicitly related to the concept of path which, in turn, depends on the routes that elements take while traversing the network. To model traffic, it is usually convenient to focus on the classical definition based on shortest path dynamics. In this context, the shortest-path betweenness ($B_n$) considers only the least costly paths (usually in length or traversal time) between city locations and is defined, for a given node $n$, as:
\begin{equation}
  B_n=\frac{1}{\mathcal{N}}\sum_{o\neq d}\frac{\sigma_{od}(n)}{\sigma_{od}},
  \label{eq:SPBdef}
\end{equation}
where $\sigma_{od}$ is the number of shortest-paths going from origin~$o$ to destination~$d$, while $\sigma_{od}(n)$ is the number of these paths crossing~$n$. Factor~$\mathcal{N}$, usually taken to be~$(N-1)(N-2)$, $N^2$, $N$ (where $N$ represents the number of nodes in the network), or even~1, represents a normalization constant which may be different depending on the application. For convenience, we will consider only the non-normalized betweenness, setting $\mathcal{N}=1$.

Despite node and edge betweenness being usually considered purely structural centrality measures, it is important to realize that, implicitly, they carry on a routing protocol, an origin-destination matrix, and a microscopic dynamics of traffic that can be tailored to adapt different situations. This allows, similar to queuing theory \cite{gawron1998iterative} or the UTA model \cite{Kerner2004}, to accurately model the expected traffic at the micro, meso and macro scale \cite{sole2016model}.

Regarding the routing protocol, recent studies indicate that drivers (and pedestrians) may opt for alternative trajectories larger than the shortest path \cite{quercia2014shortest}, but the assumption of shortest path dynamics is still an outstanding routing model to analyze. It is based on the rational choice of trajectories and, undoubtedly, has been helpful in the design and analysis of transportation networks. Other routing protocols \cite{estrada2012physics} (e.g., random walks \cite{Newman2005}) or origin-destination matrices (e.g., those based on experimental data \cite{sole2018decongestion}) lead to alternative definitions of betweenness, usually wrapped under the name of effective betweenness \cite{Guimera2002}.

As already evidenced by Guimer\`a et al.~\cite{Guimera2002}, the analysis of the betweenness distribution is crucial for the understanding of the dynamical properties of transportation networks, since it constitutes an accurate proxy of node and link usage which, combined with other properties (e.g., processing capacity), can accurately predict congestion.

\begin{figure}[t!]
  \begin{center}
  \includegraphics[width=0.98\columnwidth]{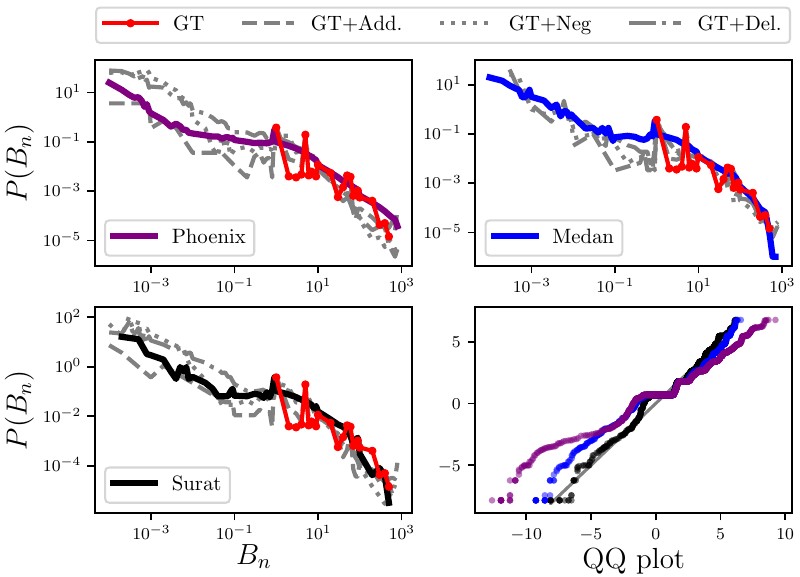}
  \end{center}
  \caption{Betweenness distribution of the road networks of three cities, Phoenix (AZ, USA), Medan (Indonesia) and Surat (India), compared with a GT-model without noise (red dotted line), and GT-model with three different kinds of noise: biased edge addition, biased edge removal, and Delaunay. The bottom right panel shows a $QQ$-plot between the real (cities) and experimental (GT-Model) betweenness distributions. The $QQ$-plot refers to the quantiles of the distribution obtained by the GT-model (vertical axis) against those related to the empirical one (horizontal axis). Similarity between the distributions improves as the QQ plot approaches the bisector, the grey solid straight line. GT-model parameters are $w=31$, $r=2$ and $h=9$, while noise has been generated until density $\rho = 0.52$ is obtained for the gravity and Delaunay type, and $\rho = 0.4$ for the negative one.}
  \label{fig:GTmodelBetwDist}
\end{figure}

Recently, Kirkley et al.\ \cite{Kirkley2018} have shown that, when networks are attributed with planar properties, such as urban road networks, these distributions display a particular shape that scales with $N$. Additionally, there is a strong dependence between node betweenness and its geographic position. In general, low betweenness nodes are located at peripheral regions, while high betweenness nodes appear near the urban center. This distribution can be seen in Fig.~\ref{fig:GTmodelBetwDist} for three different cities (see Fig.~2 of \cite{Kirkley2018} for a large scale analysis). The right hand side of the betweenness distribution can be approximated with a power-law with an exponential cut-off: $P(B_n)=B_n^{-\alpha}e^{B_n/\beta}$. Exponent $\alpha$ happens to be quite stable, with values $\alpha \approx 1$. However, the value of $\beta$ has a strong dependence on the size of the city \cite{Kirkley2018}.

\section{Grid-tree model for monocentric city road networks}
\label{sec:GTmodel}

The dependence of the shape of the tail of the betweenness distribution with respect to city size suggests that there might be a structural transition between different network topologies as we depart from the city center. Different network arrangements may coexist \cite{marshall2004streets,han2020classification}. At large distances, we mainly expect to find arterial roads \cite{marshall2004streets,Strano2012}, whose connectivity may resemble a tree \cite{marshall2004streets} (or a dendritic structure). The betweenness distribution of trees follows a power-law which is compatible with the low influence on the cut-off parameter $\beta$ on $P(B_n)$ (see left panel in Fig.~S1 of the Supplemental Material \cite{SM}). As we approach to the city center, the presence of local roads increases and the structure of the network increments in regularity \cite{kostof1991city,masucci2013,marshall2004streets}. Although many types of regular networks (e.g., Delaunay triangulations or rectangular grids) could mimic this structure, most of them have similar properties in terms of betweenness. They provide high inter-connectivity between city center buildings, with high redundancy of paths, offering, in turn, higher resilience to congestion than tree-like structures. A grid graph displays a betweenness distribution with a cut-off (see Fig.~S1 of the Supplemental Material \cite{SM}).

\begin{figure}[t!]
  \begin{center}
  \includegraphics[width=0.98\columnwidth]{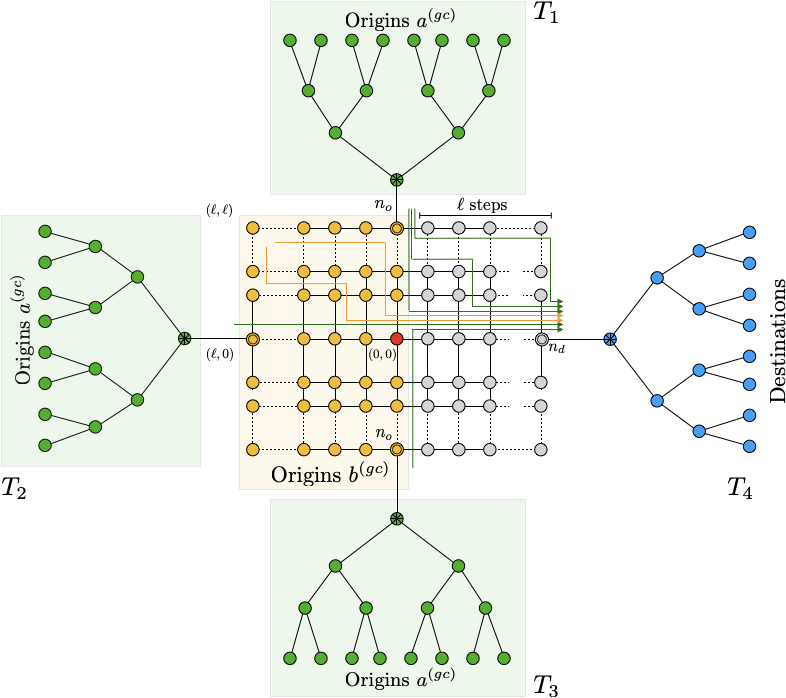}
  \end{center}
  \caption{Diagram of a network generated with the GT-model, with parameters $w=2\ell + 1$, $r=2$ and $h=3$. The maximum betweenness can be located at the grid-center node (in red), the connector nodes (marked with a double circle), or the tree-root nodes (marked with a star). Colors, labels and notation are set to explain the analytical computation of the betweenness of the central node of the grid, see Sec.~\ref{sec:AnBeetwCalc}. }
  \label{fig:GTModel}
\end{figure}

These observations allow us to set the basis of our models for road networks of monocentric cities: a dense regular planar graph in the city center embraced by a tree-like (or arboreal) network structure. In this section, city center is described by a square grid, connected to trees that model the periphery. More precisely, our \emph{grid-tree model} (simply GT-model hereafter) consists of a regular grid, of size $w\times w$, connected with a set of $n_t$ trees with height $h$ and branching factor $r$, see Fig.~\ref{fig:GTModel}. We suppose the trees are full and complete. We explore an alternative model in Sec.~\ref{sec:RandomPlanarModels}.

The number of nodes of the grid, $N_{G}$, and of the trees, $N_{T}$, are given by
\begin{equation}
  N_{ G}=w^2,\quad N_{T}=\sum_{v=0}^h r^v=\frac{r^{h+1}-1}{r-1}\,,
  \label{eq:sizes}
\end{equation}
while the size of the GT-model structure is
\begin{equation}
  N_{GT}=N_{G}+n_{t}N_{T}\,.
\end{equation}
For simplicity and symmetry of the GT-model, we choose four trees ($n_t=4$), an odd number for the sides of the grid ($w=2\ell+1$, with $\ell\in\mathbb{N}$), and assume that trees are connected to the central node of each side, see Fig.~\ref{fig:GTModel}.

Compared to actual cities of diverse size, the GT-model reproduces the range $[1, \infty)$ (right branch) of the betweenness distribution in Fig.~\ref{fig:GTmodelBetwDist} (red line in panels a-c). These are the nodes with larger betweenness centrality, the critical ones to many network phenomena. This is a clear evidence that our simple regular network model may suffice for the analysis of these phenomena. The left hand side of the rescaled betweenness distribution, range $[0, 1)$, corresponding to low values of betweenness, emerges with the addition of structural noise to the GT-model. We have tested three types of noise: random addition and removal of edges (both with distance bias), and Delaunay noise. See Appendix~\ref{sec:noise} for a detailed description of noise generation procedures.

The matching between the empirical and GT-model distribution is also portrayed through a Quantile-Quantile ($QQ$) plot, which allows to compare two distributions by plotting their quantiles against each other. The $QQ$-plots for the three cities show that the GT-model leads to similar shapes of the betweenness distribution, and especially, a good fit for the upper part of the distribution. The lower part of the distribution, corresponding to the structural noise, is slightly underestimated.

\begin{figure}[t!]
  \begin{center}
  \includegraphics[width=0.98\columnwidth]{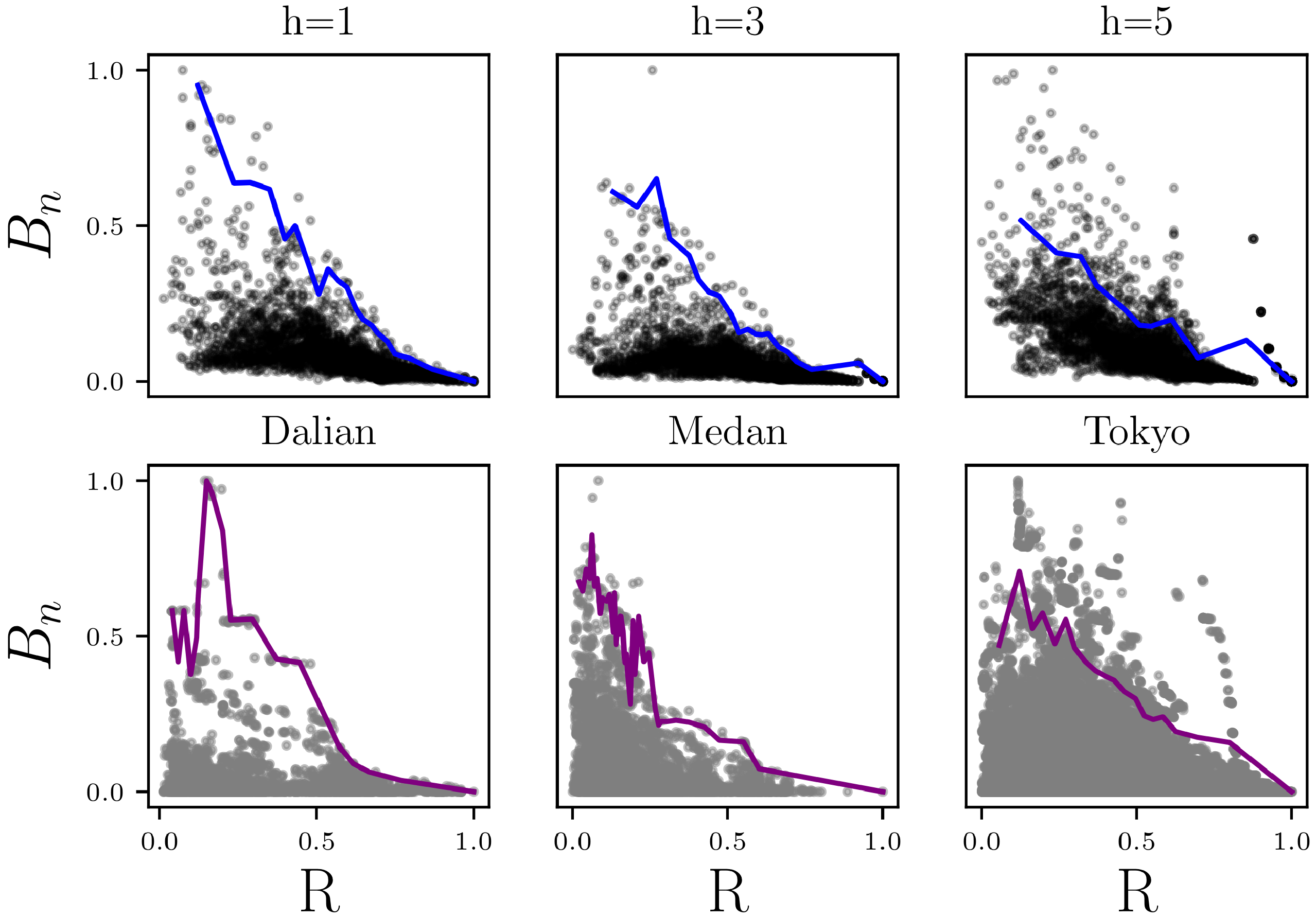}
  \end{center}
  \caption{Comparison between the real (bottom) and estimated (top) relationship between nodes geographic position and their betweenness for small (Dalian, China), medium (Medan, Indonesia) and big cities (Tokyo, Japan). The GT-models used have parameters $w=51$, $r=2$ and the values of $h$ indicated in each column. Additive noise is also added in terms of relative increments of network density, $\Delta\rho/\rho=1.5\%$ (see Appendix~\ref{sec:noiseadditive}). All distances and betweenness are normalized between 0 and 1. Coordinates for the nodes of the GT-model have been assigned using the planar embedding described in Appendix~\ref{sec:PlanEmbed}. Solid lines on the scatter plots represent the 80-centiles of the betweenness distribution at the given radius $R$; similar results for the mean and standard deviations are provided in Fig.~S2 of the Supplemental Material \cite{SM}).}
  \label{fig:GTmodelSPBdistMax}
\end{figure}

Additionally, as important as the shape of the distribution, is the relationship between the node's geographic position and its betweenness. We show in Fig.~\ref{fig:GTmodelSPBdistMax} that the GT-model recovers the monotonic decrease of the larger betweenness nodes, as well as the average betweenness, as we move away from the city center (see Fig.~S2 of the Supplemental Material \cite{SM}). Furthermore, it not only recovers the general trend but also the deviation of these values, i.e., large deviations for nodes near the city center, and low deviations for peripheral regions (see Fig.~S2 of the Supplemental Material \cite{SM}).

These positive results depend, of course, on the choice of the parameters. The accuracy of the GT-model is determined by the values taken by $w$, $r$, $h$, and the level of noise. In particular, it is governed by the ratio between $N_{G}$ and $N_{T}$. As this ratio grows, the high-betweenness distribution branch approaches the form of a grid and the cut-off~$\beta$ has an influential role, while in the opposite case a power-law emerges as it is related to the betweenness distribution of the tree. Results in Fig.~\ref{fig:GTmodelSPBdistMax} correspond to intermediate regimes. They show a theoretical-empirical comparison of the betweenness spatial behavior. For this objective, we have selected small (Dalian), medium (Medan) and large (Tokyo) cities, and compared them with the GT-model generated with different tree sizes, namely with $h=1,3,5$. We have chosen to extend the trees instead of the grid because as cities grow (see Sec.~\ref{sec:BeetwDist}) the tree structure becomes predominant. We present the betweenness spatial behavior for grid and trees in Fig.~S3 of the Supplemental Material \cite{SM}.

\section{Analytical derivation of betweenness in the GT-model}
\label{sec:AnBeetwCalc}

For a large set of traffic models and strategies, the critical injection rate of vehicles, $\gamma_c$, i.e., the maximum rate at which vehicles can enter the system without congesting it, can be obtained in terms of the maximum node betweenness, $B_n^{\ast}$ \cite{Guimera2002,zhao2005onset,Echenique2005,yan2006efficient,liu2007method,dong2012enhancing,tan2014traffic,sole2016congestion,manfredi2018mobility}:
\begin{equation}\label{CongParam}
  \gamma_c = \frac{N_{GT}-1}{B_n^{\ast}}\,.
\end{equation}
The value of $B_n^{\ast}$ sets the onset of congestion. Thus, it makes sense to study routing dynamics over the GT-model in such terms. Here, we are concerned about how topological changes in terms of the GT-model model parameters (i.e., $w$, $r$, $h$ and noise), may affect the position of $B_n^{\ast}$ in the network.

Numerical exploration of the (noiseless) GT-model reveals that the onset of congestion is set by nodes located at three different key network positions: the center node of the grid (red node in Fig.~\ref{fig:GTModel}), at connector nodes on the perimeter of the grid (nodes with a circumscribed circle in Fig.~\ref{fig:GTModel}), or at the root of the trees (nodes with a star within in Fig.~\ref{fig:GTModel}). We will refer to these nodes as grid-center~(gc), connector~(c), and tree-root~(t), respectively. According to them, it is possible to define three different congestion regimes, which correspond to the location of the node that marks the onset of congestion. This indicates that the transportation network may collapse in the city center, in the perimetral city roads (ring roads), or in arterial roads, unveiling a new spatio-dynamical property of network congestion. Note that in these two last cases, because of symmetry, we have four nodes with equivalent structural properties (see Fig.~\ref{fig:GTModel}).

According to Eq.~(\ref{CongParam}), the understanding of which circumstances lead to each regime, goes through the analysis of the betweenness for these three types of nodes. The rest of this section develops the analytical expressions of the betweenness for these nodes of interest, while we leave for the next Sec.~\ref{sec:CongestionRegimes} the analysis of the congestion regimes. To this aim, we proceed in the classical way of counting the number of paths crossing each of the nodes, according the definition in Eq.~(\ref{eq:SPBdef}). As we will see, regularities in the network allow for an efficient computation of such values. For the sake of simplicity, we assume that origin and destination nodes do not contribute to the betweenness of such nodes. Otherwise, one should add $2(N_{GT}-1)$ to each node: $N_{GT}-1$ for the paths where the node is origin, and $N_{ GT}-1$ for the paths where the node is destination.

Considering the structural composition of the GT-model, each node may be traversed by three different types of paths: (1) paths with origin and destination in the trees; (2) paths between a tree and the grid; (3) paths within the grid. The betweenness of each node can be obtained as the sum of the contributions of each of these types of paths. Specifically, the betweenness of any node in the network can be written as a second-degree polynomial:
\begin{equation}
  B_{j}=a_j N^2_{T} + b_j N_{T} + c_j\,,
  \label{eq:BmaxGTModel}
\end{equation}
with different coefficients depending on the GT-model parameters. The constant term $c_{j}$ considers the contribution of paths fully contained within the grid, which does not depend on the parameters of the tree. The linear term $b_{j} N_{T}$, instead, considers the contribution of paths that go between a tree and the grid, while the first term, $a_{j} N^2_{T}$, considers the contribution of paths that go between trees. All the coefficients $a_{j}$, $b_{j}$, and $c_{j}$ depend on the grid parameter ($w$ or $\ell$), and for the tree-root node, also on~$r$.

The analytical computation of the coefficients of $B_j$ is cumbersome. Consequently, to improve readability  of the paper, in the following, we only describe the mechanism we use to obtain the betweenness for the grid-center node ($B^{(gc)}$), and postpone the derivation of the betweenness for the connector node ($B^{(c)}$) and the tree-root node ($B^{(t)}$) to Appendices~\ref{sec:Connector} and~\ref{sec:FullCompleteTree}, respectively.

Coefficient $a^{(gc)}$ of the betweenness of the grid-center node can be expressed as:
\begin{equation}
  a^{(gc)} = 2 + 4\frac{1}{\pi_{\ell,\ell}}
  = 2 + 4\frac{(\ell!)^2}{(2\ell)!}\,,
  \label{Eq:a_gc}
\end{equation}
where $\ell=(w-1)/2$ corresponds to the distance between the central node of the grid and its sides, see Fig.~\ref{fig:GTModel}, and we have defined $\pi_{x,y}$, the number of different paths in the grid involving $x$~horizontal and $y$~vertical steps:
\begin{equation}
  \pi_{x,y} = \binom{x + y}{x} = \frac{(x + y)!}{x!\,y!}\,.
\end{equation}
As described in Eq.~(\ref{eq:BmaxGTModel}), $a^{(gc)}$ only considers the contribution to betweenness of paths with origin and destination belonging to nodes in different trees. Following Fig.~\ref{fig:GTModel}, which visually describes the process, we first consider the betweenness contribution to the grid-center node (colored in red) of paths that go from $T_2$ to $T_4$. Each of these paths contributes with a unity to $B^{(gc)}$, since all paths between those nodes go though the central node. Taking into account also the paths between $T_1$ and $T_3$, we obtain the factor~$2$ in Eq.~(\ref{Eq:a_gc}).

Now consider the contribution of paths that go from either $T_1$ or $T_3$ to nodes within $T_4$. All these paths are canalized through nodes $n_o$ and $n_d$, and there are many paths of equal length between $n_o$ and $n_d$: we have path multiplicity (or degeneration). Some of them are illustrated in green lines in Fig.~\ref{fig:GTModel}. We proceed combinatorially to count all these paths. Consider we need to move $\ell$ steps to the right ($\rightarrow$) and $\ell$ steps down ($\downarrow$) to go from $n_o$ to $n_d$. There are $\pi_{\ell,\ell} =\binom{2l}{l}$ ways in which we could order the $\rightarrow$ and $\downarrow$ operations. Only one of these paths goes through the central node, the one where all $\downarrow$ operation precede the $\rightarrow$ ones. In this way, the betweenness contribution of any of the paths that go from nodes in $T_1$ or $T_3$ to nodes within $T_4$ is $1/\pi_{\ell,\ell}$. Considering there are four different origin-destination combinations in this configuration ---($T_1,T_4$), ($T_3,T_4$), ($T_1,T_2$) and ($T_3,T_2$)--- we add a factor~4 to Eq.~(\ref{Eq:a_gc}). Note that we do not have to account for the reversed assignment of the trees as origin and destination, e.g., ($T_4,T_1$), due to the reversibility of the paths; if we calculate all the shortest paths from node~$i$ to node~$j$, it is not necessary to do the same for paths between~$j$ and~$i$.

The expression for $b^{(gc)}$ is the following:
\begin{align}
  b^{(gc)}
  &= 4\ell + 8\sum_{x=0}^{\ell}\sum_{y=1}^{\ell} \frac{\pi_{x,y}}{\pi_{x+\ell,y}}  \nonumber
  \\
  &= 4\ell + 8\sum_{x=0}^{\ell}\sum_{y=1}^{\ell} \frac{(x+y)!\,(x+\ell)!}{x!\,(x+y+\ell)!}.
    \label{Eq:b_gc}
\end{align}
Figure~\ref{fig:GTModel} provides a visual support for the explanation of the terms in Eq.~(\ref{Eq:b_gc}). Consider a node identified with variables $(x,y)$, i.e., located at $x$~horizontal and $y$~vertical steps of the central node of the grid. The shortest paths whose destination node is within tree $T_4$ and that go through the central node, are paths whose origin is located on the left-hand side of the grid, shaded in orange in the diagram. Any of these origin-destination pairs has $\pi_{x,y}$ different paths that cross the central node, and $\pi_{x+\ell,y}$ paths to reach the connector node that connects the grid and the tree, since the connector node is $\ell$ nodes to the right of the central node. Thus, each of the origin-destination pairs contribute to the betweenness of the central node with a factor $\pi_{x,y} / \pi_{x+\ell,y}$.

We can now exploit the grid's symmetries to obtain their total contribution to the betweenness of the central node. For each origin node above the central node there is an equivalent node below it, thus a factor~2 must be added. However, for nodes with $x=0$, there is just one path to the destination, and it crosses the central node, thus the term~$\ell$ in front of the sums. Finally, a factor~4 to both terms is necessary to account for the 4~possible destination trees connected to the grid, thus completing all the terms in Eq.~(\ref{Eq:b_gc}).

The calculation of coefficient $c^{(gc)}$ is similar to the previous ones, with just the difference that both origin and destination of the paths belong to the grid. The result is
\begin{align}
  c^{(gc)}
  =\ &2 \ell^2
    + 4 \sum_{a=1}^{\ell} \sum_{y=1}^{\ell} \frac{1}{\pi_{a,y}}
    + 8 \sum_{a=1}^{\ell} \sum_{x=1}^{\ell} \sum_{y=1}^{\ell} \frac{\pi_{x,y}}{\pi_{x+a,y}} \nonumber
  \\
    &+ 2 \sum_{a=1}^{\ell} \sum_{b=1}^{\ell} \sum_{x=1}^{\ell} \sum_{y=1}^{\ell} \frac{\pi_{x,y}\,\pi_{a,b}}{\pi_{x+a,y+b}}\,.
    \label{Eq:c_gc}
\end{align}
The idea is to consider that the origin node is identified with variables $(x,y)$ and the destination node with $(a,b)$, where~$x$ and~$a$ represent the horizontal distances to the grid center, while~$y$ and~$b$ are the corresponding vertical distances. There exist four different configurations for the relative positions of these origin and destination nodes, that lead to the four terms in Eq.~(\ref{Eq:c_gc}).

In the first term, the origin and destination are aligned with the grid center, i.e., $x=a=0$ for a vertical alignment, and $y=b=0$ for the horizontal alignment. They amount a total of $\ell^2$~non-degenerated shortest paths per alignment, in which all of them contain the grid-center node.

The second term considers the cases in which origin and destination nodes are, one aligned horizontally and the other vertically with the grid center, i.e., the cases $x=b=0$ and $y=a=0$. There are four of these combinations, and in all of them there is only one shortest path that crosses the grid center. If we choose for example the case $x=b=0$, there exist a total of $\pi_{a,y}$ shortest paths between the origin and the destination, thus explaining this term in Eq.~(\ref{Eq:c_gc}).

Next, we have the situation in which the alignment with the grid center is limited to one of the nodes, and in one of the directions; let us choose the destination node and the horizontal alignment, i.e., $b=0$. Now, only $\pi_{x,y}$ of the $\pi_{x+a,y}$ shortest paths connecting origin and destination pass through the grid center. Since there are eight possible origin-destination configurations with this relative alignment, they explain the third term in the r.h.s.\ of Eq.~(\ref{Eq:c_gc}).

Finally, the last term comes from the two configurations where no alignment is present, with $\pi_{x,y}\,\pi_{a,b}$ shortest paths crossing the grid center among the $\pi_{x+a,y+b}$ connecting origin and destination.

Once we have computed the three coefficients, the betweenness of the grid center is simply:
\begin{equation}
  B^{(gc)} =a ^{(gc)} N^2_{T} + b^{(gc)} N_{T} + c^{(gc)}\,.
  \label{eq:BmaxGC}
\end{equation}
Note that the same approach could have been used to obtain the betweenness of all the nodes in the grid, not just the grid center. Basically, the limits of the sums in Eqs.~(\ref{Eq:b_gc}) and~(\ref{Eq:c_gc}) would change to reflect the new position of the reference node, some terms would disappear from the coefficients due to the node not being within a shortest path, and special care would be needed to account for nodes in the sides and corners of the grid. The analysis for one type of these nodes, the connector node, is available in Appendix~\ref{sec:Connector}, while in Appendix~\ref{sec:FullCompleteTree} we show the calculation of the betweenness for all of the nodes in the trees, including the important tree-root node.

\begin{figure}[t!]
  \begin{center}
  \includegraphics[width=0.98\columnwidth]{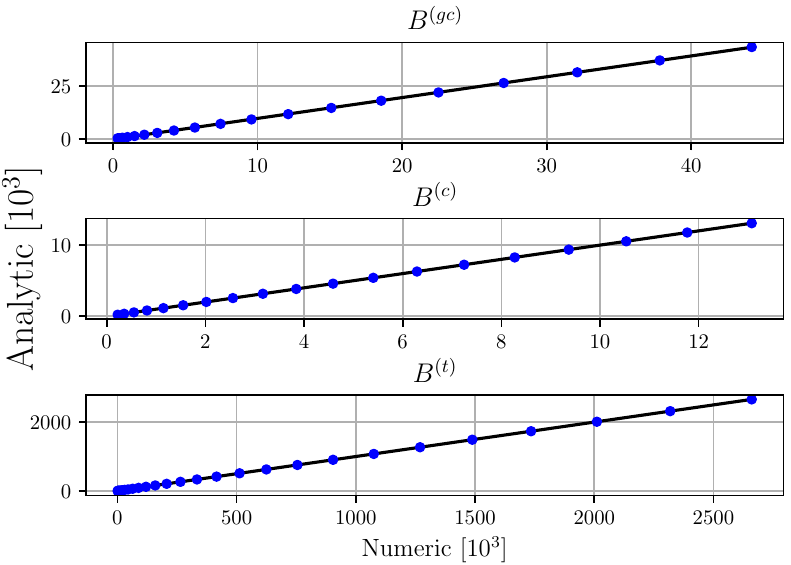}
  \end{center}
  \caption{Comparison between the analytical and numerical betweenness for the grid's central node (top), connector node (middle), and tree-root node (bottom) of the GT-model. Analytical values are obtained by means of Eq.~(\ref{eq:BmaxGTModel}), while numerical ones are calculated using the Brandes algorithm \cite{brandes2001faster}. Both $B^{(gc)}$ and $B^{(c)}$ are obtained by varying $w\in[3,40]$ at $r=h=2$, while $B^{(t)}$ is treated as a function of $r\in[3,30]$ at fixed $h=2$ and $w=5$.}
  \label{fig:GTmodelBmax}
\end{figure}

For validation purposes, Fig.~\ref{fig:GTmodelBmax} displays a correlation plot between the analytical values of the betweenness calculated using the expressions above, and the numerical values obtained using the Brandes algorithm \cite{brandes2001faster}, showing perfect agreement.

\section{Congestion Regimes}
\label{sec:CongestionRegimes}

The position of the maximum betweenness node provides information about which city areas determine the collapse of the transportation network. The three regimes (central's grid, connector and tree-root node) we have discussed in the previous sections may be interpreted as different congestion phases which characterize an urban system. In this section, we establish a relation between the three congestion regimes and the parameters of the GT-model. Precisely, given the values of $w$, $r$ and $h$, we conclude where, geographically, congestion occurs.

The transition between two different regions is defined by the condition
\begin{equation}
  B^{(r_1)}-B^{(r_2)}=0,\quad r_1\neq r_2,
\end{equation}
where $r_1$ and $r_2$ are two distinct regimes. For instance, the frontier between the tree-root ($t$) and the connector node ($c$) regimes is defined by the equation
\begin{equation}\label{TrGsFront}
  B^{(t)}-B^{(c)}=0,
\end{equation}
with $B^{(t)}$ and $B^{(c)}$ introduced in Eq.~(\ref{eq:BmaxGTModel}). Clearly, Eq.~(\ref{TrGsFront}) is a second-degree polynomial:
\begin{equation}\label{TrGsFrontPol}
  a^{(t,c)}N^2_T+b^{(t,c)}N_T+c^{(t,c)}=0
\end{equation}
with
\begin{align}
  a^{(t,c)} &= a^{(t)}-a^{(c)},
  \\
  b^{(t,c)} &= b^{(t)}-b^{(c)},
  \\
  c^{(t,c)} &= c^{(t)}-c^{(c)}.
\end{align}
Then, by using the equations in Sec.~\ref{sec:AnBeetwCalc}, and Appendices~\ref{sec:Connector} and~\ref{sec:FullCompleteTree}, we can provide an analytic solution to Eq.~(\ref{TrGsFrontPol}):
\begin{equation}\label{eq:NumTAn}
  N_T=\frac{-b^{(t,c)}+\sqrt{\Delta^{(t,c)}}}{2a^{(t,c)}},
\end{equation}
with the discriminant
\begin{equation}
  \Delta^{(t,c)}=\left(b^{(t,c)}\right)^2-4a^{(t,c)}c^{(t,c)}.
\end{equation}
In a similar way we can obtain the transition between the other regimes.

\begin{figure*}[t!]
  \begin{center}
  \includegraphics[width=0.98\textwidth]{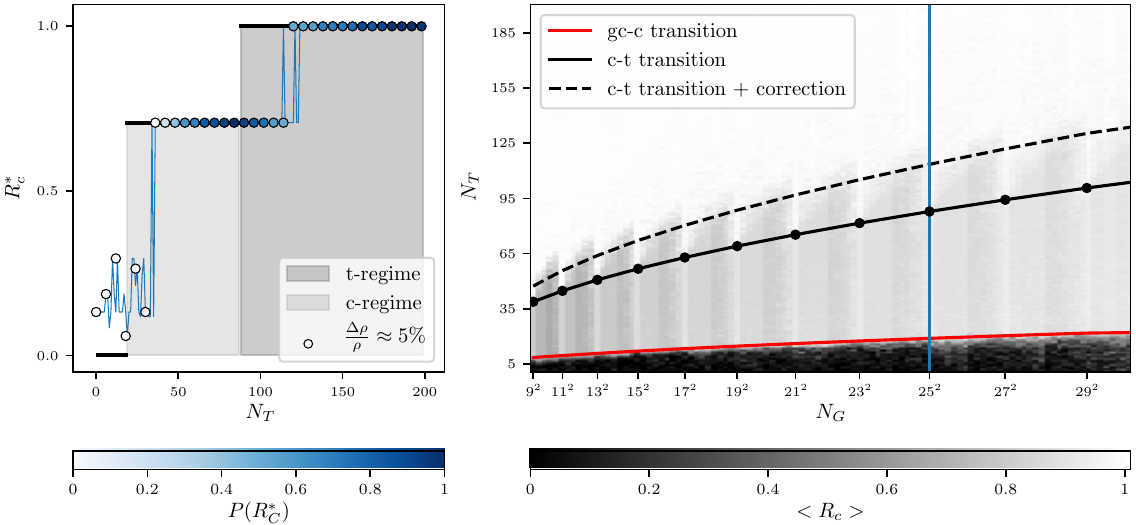}
  \end{center}
  \caption{Congestion phase space of the GT-model, showing the three regimes and their transitions. Left: Solid black lines refer to the regimes as predicted by Eq.~(\ref{eq:NumTAn}), with $w=25$ and $r=2$. Circles represent the experimental results after the addition of noise. Each circle is located at the statistical mode (here indicated by $R^{*}_c$) obtained with the distribution of $R_c$ after 150~realizations. The color of the circle shows the probability of that value over the experimental $R_c$ distribution. See Figs.~S4 and ~S5 of the Supplemental Material \cite{SM} for an equivalent analysis with lower and higher levels of noise ($\Delta \rho / \rho = 0.2\%$ and $23\%$, respectively), which show that the transitions persist even for high levels of noise, and their shift to the right vanishes with low noise. Right: Average distance, in GT-model, between the maximum betweenness node and the grid center as a function of the number of nodes of the grid $N_G$ and the trees $N_T$, for a fixed value of the branch parameter $r=2$. Average is carried out over 20~model configurations with noise $\Delta\rho/\rho = 0.2\%$. The result is normalized considering the maximum obtained over the set of configurations related to the same grid size ($N_G$). Red line corresponds to the estimated frontier between the grid-center and the connector regime, while black lines (solid and dashed) correspond to the transition between the connector regime and tree-root one, as predicted by Eq.~(\ref{eq:NumTAn}). Blue vertical line highlights the system at $N_G=w^2=25^2$  and corresponds to the results of the left panel.}
  \label{fig:PhaseDiagram}
\end{figure*}

Figure~\ref{fig:PhaseDiagram} shows the three different regimes for varying configurations of the GT-model. When the trees are small with respect to the grid, the grid-center node dominates congestion. Increasing the size of the trees, the congestion jumps first to the connector nodes, and later to the tree-root nodes. An appropriate parameter to know the current congestion regime is the congestion radius~$R_c$, defined as the distance between the maximum betweenness node and the grid center.

Once noise is added to the GT-model using the methods in Appendix~\ref{sec:noise}, the regime's transitions are expected to soften. The left panel of Fig.~\ref{fig:PhaseDiagram} presents the behavior of the congestion radius as a function of $N_T$ for fixed~$r$ and~$w$. First, for small sizes of the trees, we observe an offset in the grid-center regime, although the congestion radius remains close to the grid center ($R_c=0$). Secondly, as tree height increases, we observe an expected noisy behavior preclusion of the transitions between the regimes. Finally, a general shift of the different transition with respect to the noiseless case is also pointed out. However, despite all, the different regimes are still clearly identifiable and stable for a wide range of~$N_T$ values when noise is added to the GT-model. It is worth highlighting that the abruptness of the transition remains despite the noise, and its slight smoothing effect.

The right panel of Fig.~\ref{fig:PhaseDiagram} generalizes the results on the left one and analyzes, in terms of a phase diagram, the accuracy in predicting the transition between the different regimes for a large set of model parameters. As defined, the GT-model only considers squared grids and regular trees, which probably is too restrictive to resemble real cities. Also, it renders sparse model sampling as the parameters get large. To overcome these drawbacks, we extend the model to incorporate intermediate grid and tree sizes (besides the ones given by Eq.~(\ref{eq:sizes}) which allows to draw GT-model realizations of any network size. In the grid case, we generate these intermediate sizes between $w^2$ and $(w+1)^2$ by incrementally adding nodes (one by one) to the current grid periphery until we reach the desired nodes number $w^2 \leqslant N_G \leqslant (w+1)^2$. In this iterative process, nodes are added at random locations considering the remaining empty positions. We also implement an equivalent procedure for the tree.

Results in the right panel of Fig.~\ref{fig:PhaseDiagram} are intuitive: congestion occurs in the center of the grid (dark values in the phase diagram) when trees are short, i.e., grid dynamics predominate. As trees increase in size (vertical axis) a phase transition to the connector regime occurs (gray values in the diagram). At this point, the connector nodes become a bottleneck for the transportation network. Eventually, as trees keep increasing in size, the highest hierarchy nodes of trees becomes the bottleneck and then we reach the tree-root regime (light gray in the diagram). These transitions from the center to the outer bounds of the graph are reminiscent of the double percolation phase transition observed in core-periphery networks~\cite{Colomer2014}. Indeed, the GT-model shares some structural features with that family of networks, which may explain the resemblance of such observation. Finally, note that the phase transitions evidenced in the left panel of Fig.~\ref{fig:PhaseDiagram} represent a vertical slice of the phase diagram in right one (marked as a blue vertical line).

We see that the transition between the grid center and the connector regimes, predicted by Eq.~(\ref{eq:NumTAn}), is accurate in all the phase diagram (red line). The transition between the connector and the tree-root regimes, as given by Eq.~(\ref{eq:NumTAn}), is only accurate at grid sizes given by Eq.~(\ref{eq:sizes}) (black solid line with dots in Fig.~\ref{fig:PhaseDiagram}). This is mainly because the addition of individual nodes to the side of the grid contributes unequally to the betweenness of the connector and the tree root. In comparison with the regular sizes of the grid (circles) we need much larger trees to trespass the phase transition. This can be considered in Eq.~(\ref{eq:NumTAn}) without much effort to obtain a better prediction (black dashed line). A detailed explanation is given in Appendix~\ref{sec:noncompletegrid}.

In addition to the detection of the phase transitions, our analytical development allows to understand the asymptotic behavior of the regimes. Here, the quadratic character of the transition function in Eq.~(\ref{TrGsFrontPol}) permits to state that the two regimes never collapse, neither reach a constant separation, and the grid-side area enlarges as $N_G$ increases. This would mean that, as cities grow larger, the internal flow predominates the dynamics of the transportation system. This is compatible with recent observations considering real transport data \cite{louail2015uncovering}.

Finally, it is interesting to analyze how the different congestion regimes are affected by the amount of trees connected to the grid. Supported by the results in Fig.~S6 of the Supplemental Material \cite{SM}, where we analyze the congestion radius as a function of the number of trees connected to the grid, we conclude that the connector and tree root regimes slowly merge as the number of trees is increased. These results are additionally validated in Fig.~S7 of the Supplemental Material \cite{SM}, where we study a particular case with a large number of trees connected to the grid. In general, the connection of further peripheral structures to the grid perimeter implies a large increase of the betweenness of the connector nodes, which get a high score regardless of the trees height.

One could also consider the situation in which the number of trees is fixed, but increasing the number of connections between their roots and the grid perimeter. This alternative situation gives high betweenness scores to the tree root, and the corresponding regime absorbs the connector one. In conclusion, the individual presence of the connector and tree root regime is influenced by the number of trees and the related entry/exit points, i.e., on how grid-tree spatial discontinuity is engineered. Nevertheless, regardless of these microscopic details, the existence of an abrupt transition between center regime and one of the remaining two is robust, see Fig.~S6 of the Supplemental Material \cite{SM}.

\section{DT-MST Random Planar Model}
\label{sec:RandomPlanarModels}

The results presented in the previous section have been obtained by means of the GT-model introduced in Sec.~\ref{sec:GTmodel}. The main advantage of this model lies in its analytical tractability, that permits to characterize the possible congestion regimes and the related abrupt transitions. However, as usual, this comes at the cost of hard assumptions and oversimplifications. The GT-model depends on a number of choices concerning its microscopic ingredients, such as the number of connectors or the trees regularity.

In the current section we relax many of these constraints, allowing us to identify the fundamental properties of street networks behind the discovered abrupt transitions. Specifically, we construct random planar networks which keep the general idea of arterial-central roads structure and entanglement, while being as free as possible from any other artificial assumption. In this framework, the GT-model constitutes a particular case in which the center and the arterial periphery can be approximated to a grid and a regular tree respectively, for analytical purposes.

To construct these random planar networks, that we call DT-MST model, we relay on a node embedding over a two-dimensional space, and two network generation algorithms: a Delaunay Triangulation (DT), and a Maximum Spanning Tree (MST). We use the network structure obtained from the DT to model the central core of the city with radius $r\leqslant R^{(DT)}\equiv\sqrt{2}\,\ell$. In this way, the high abundance of intersections, constituting a distinguishing feature of urban roads, arises automatically. Along the same line, the peripheral tree-like structure is constructed by means of the Maximum Spanning Tree (MST) obtained from the DT edges. It is important to underline that MST is constructed maximizing with respect to edge betweenness, rather than minimizing euclidean distances. This is crucial to obtain the main routing structure that jointly considers network flow and distance.

Since we are interested in progressively relaxing the assumptions imposed by the GT-model, we introduce a new parameter, $\Delta N/N\in [0,1]$, to indicate the amount of nodes that are randomly displaced from the original node embedding of the GT-model (as shown in Fig.~\ref{fig:planar} of Appendix \ref{sec:PlanEmbed}), with fixed values of $w$, $L$, and $r$. The limit $\Delta N/N=0$ corresponds exactly to the positions of the GT-model, while $\Delta N/N=1$ defines the situation in which all node positions are distributed uniformly at random. Once we have the positions of the nodes, we perform a DT on them to obtain a network. Next, we calculate the betweenness of all the edges, and use them to build its MST. Finally, the network corresponding to this DT-MST model is formed by the links of the DT inside the region $r\leqslant R^{DT}$ (the center area), while its periphery is described by the links of the MST not included in the DT. Note that, for $\Delta N/N=0$, the DT-MST model recovers a similar structure than the GT-model, although not exactly the same (e.g., the quadrangular grid has been replaced by a DT).

\begin{figure*}[tp!]
  \begin{center}
  \includegraphics[width=0.99\textwidth]{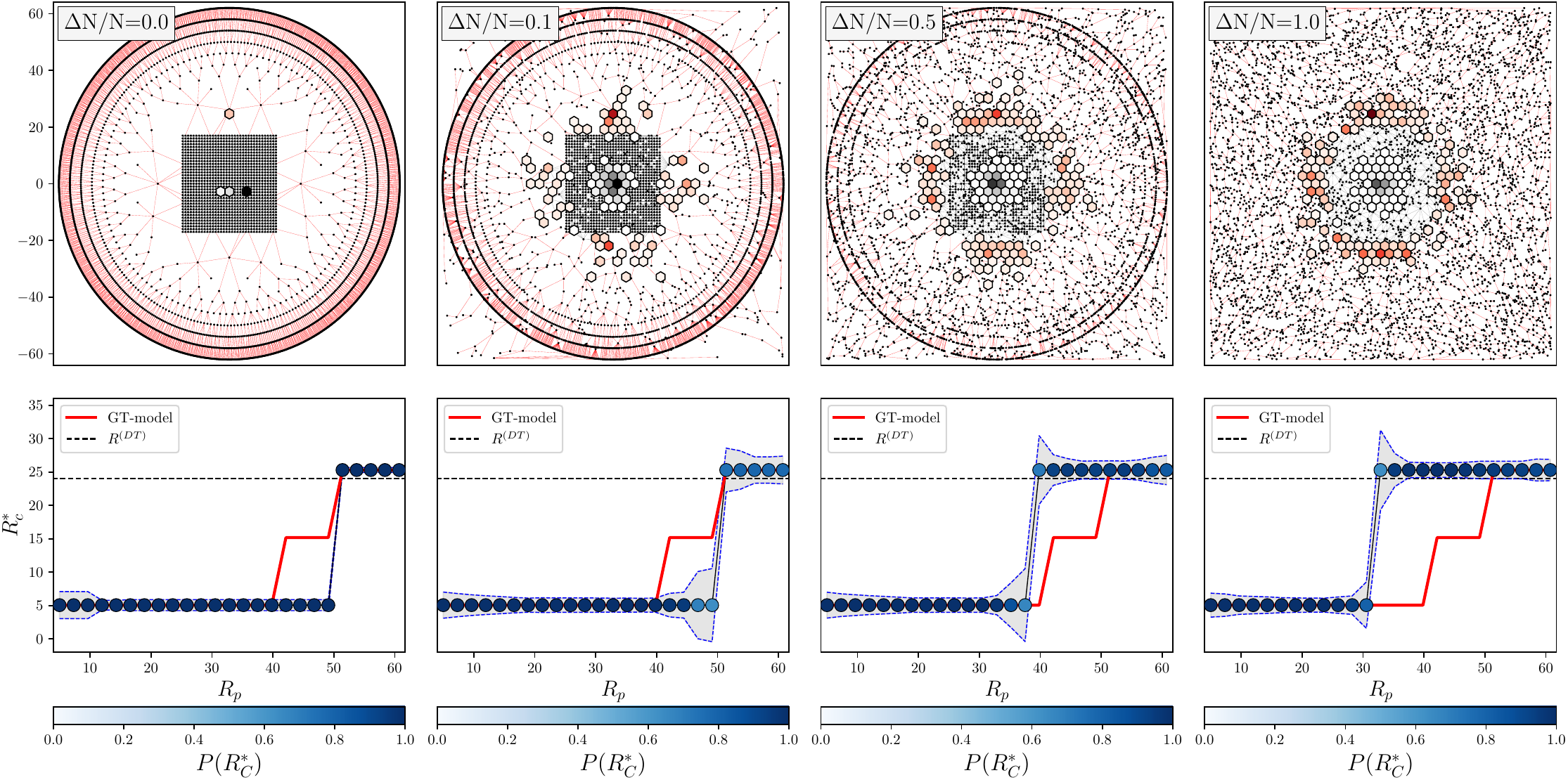}
  \end{center}
  \caption{Spatial behavior of congestion nodes for the random planar models introduced in Sec.~\ref{sec:RandomPlanarModels}, generated from a GT-model realization with $w=35$, $h=9$, and $r=2$. Each column corresponds to a different amount of the spatial noise, i.e., to different values of $\Delta N/N$. In the upper row, the resulting network configurations are shown, with MST and DT edges painted in red and black, respectively. Hexagonal bins provide information about the occurrence of congestion nodes in space, and their colors show the related frequency in proportion to the intensity. In the bottom row, we present the dependence of the congestion radius $R_c$ on the patch radial size $R_p$, i.e., the radius of the circle centered in the middle of the the network, that is used to define the considered subgraph within this radius. The congestion radius range is divided in bins, and each point is located at the statistical mode $R_c^{\ast}$ obtained with the distribution of $R_c$ after 100~realizations, as in Fig.~\ref{fig:PhaseDiagram}. The color of the circular markers shows the probability of that value over the experimental $R_c$ distribution. Shadow areas represent the variance of the different realizations of $R_c$ values with respect to the bin average value. The red solid line represents the GT-model as presented in Sec.~\ref{sec:GTmodel}, without any noise addition.}
  \label{fig:RandomPlanarModels}
\end{figure*}

Figure~\ref{fig:RandomPlanarModels} shows the resulting network configurations and its congestion analysis for $\Delta N/N=0.0,0.1, 0.5,1.0$. Clearly, for $\Delta N/N=1$, all dependence on the GT-model microscopic features are lost. First of all, nodes positions follow a random uniform distribution, and both the grid and trees symmetries are lost. The latter are not regular anymore, namely their branching factors are no longer fixed to a constant value. In general, the number of trees and connectors of the DT-MST model, as well as their positions, follow from the MST construction and may take any value. As opposed to the GT-model, the center area is circular.

Hexagonal bins in Fig.~\ref{fig:RandomPlanarModels} provide the information about the location of maximum betweenness nodes, and portray a very clear behavior: congestion either occurs in the center, or in the connection of the center with the periphery. This behavior can also be inferred from the bottom panels, where we study the congestion radius as a function of the patch radius. Note that this procedure is equivalent to the one followed in Sec.~\ref{sec:CongestionRegimes}, in which congestion radius is analyzed as a function of the tree size. Similarly to the analysis provided in Fig.~\ref{fig:PhaseDiagram}, we recover an abrupt transition between the congestion regime characterized by congestion appearing in the center and the one where congestion happens in the tree-root, proving that its existence does not depend on the specific assumptions underlying the construction of the GT-model.

The merging of the connector and tree-root regimes can be explained by the high number of trees connected to the DT region, which follows automatically from the MST construction. This fits well the discussion at the end of Sec.~\ref{sec:CongestionRegimes}, where we have analyzed the effect of having an increasing number of trees connected to the grid in the GT-model. Therefore, we may conclude that an abrupt transition between the connector and tree-root regimes exists provided the number of trees is small enough; otherwise, these regimes are merged into just one effective regime.

To further analyze the robustness of the transition between center and periphery, in Fig.~S8 of the Supplemental Material \cite{SM} we provide results for different values of $R^{(DT)}$ (DT size), which show that the transition also arises for different relative sizes of the DT and the MST regions. Remarkably, transitions are dropped out when the network is only composed by the DT or the MST, see Fig.~S9 of the Supplemental Material \cite{SM}. In these situations, congestion only occurs in the center, independently of the patch size.

All the previous analysis indicate that the spatial separation seems to be crucial for the emergence of the transition. To finally confirm this, we have modified the model to generate networks where we increasingly overlap random DT edges on an underlying MST in the same spatial region. Results in Fig.~S10 of the Supplemental Material \cite{SM} show that homogeneously mixing such structures prevents the appearance of transitions.

Summarizing, we have shown that the fundamental ingredient causing the emergence of an abrupt transition between the center and the periphery is the interplay of the two different structures composing the network, a peripheral tree-like structure and a regular and dense center. Specifically, it depends on the existence of structural discontinuities between the networks composing the system. Additionally, the second abrupt transition between the connector and tree regimes requires the existence of a low number of trees and connector nodes, otherwise these regimes merge.

\section{Evidence of Congestion Regimes in Real Cities}
\label{sec:RealCities}

The congestion regimes and the abrupt transitions uncovered by our theoretical models have not been previously observed in real cities. In this section, we empirically assess for a large set of cities, the existence of these regimes and study how abrupt their transitions are.

Specifically, we center our analysis in the structure of each city's road network contained within different concentric circles of radius $R_p$, using as geometric center of the city the one provided by an online service \cite{latlong}. This is equivalent to the procedure we have implemented in Sec.~\ref{sec:RandomPlanarModels}.
The $R_p$ radius has a clear urban interpretation: it defines the city portion where the traffic dynamics is concentrated, and out of which it is supposed to be negligible. So, one may tune $R_p$ in order to approach different mobility contexts: large values describe situations where arterial roads and traffic flows from/to city outskirts (or alternatively dormitory cities) are significant, while low values describes situations where only within city traffic is significant.

For each value of $R_p$, we obtain the set of $\lambda$ nodes with largest betweenness, and calculate the average congestion radius as
\begin{equation}\label{eq:RLambda}
  \bar{R}^{(\lambda)}_{c}=\frac{1}{\lambda}\sum^{\lambda}_{i=1}R_i\,,
\end{equation}
the average distance to the center of the city of these $\lambda$-largest betweenness nodes. In the study of the synthetic networks (GT and DT-MST models), we limited the analysis to the largest betweenness node ($\lambda = 1$). For the GT-model, its location was exactly calculated analytically. In the case of the DT-MST model, the robustness of the results was accomplished by averaging over different realizations of the model, thus ensuring their statistical significance. For real urban road networks, the statistical significance is achieved by using $\lambda>1$; in particular, from now on we make use of $\lambda=15$, and leave the study of the dependence on this parameter to the Supplemental Material \cite{SM}.

The theoretical analysis for the identification of the different congestion regimes, presented in Secs.~\ref{sec:CongestionRegimes} and~\ref{sec:RandomPlanarModels}, is based on the variation of the congestion radius with respect to the patch radius. Both network models, GT-model and DT-MST model, show a clear distinction between the different regimes. However, real cities present more challenging scenarios where the structural separation between the city center and its periphery may not be as clear. To overcome this issue without falling into arbitrary assumptions, we have designed an automated and unsupervised approach to statistically find the congestion regimes in real road networks. The method is based on identifying change points where the congestion radius significantly changes, either in mean or slope \cite{killick2012optimal}. Additionally, we make use of the elbow method (usually applied to k-means clustering algorithm) to choose the optimal number of change points, see Appendix~\ref{sec:elbow}. Subsequently, regions between two consecutive change points will define the congestion regimes.

We have applied our analysis to the 97~city road networks in \cite{Kirkley2018}, using the data provided by the authors; a detailed description the data can be read in the reference. Overall results show that 52~networks present clear, detectable at naked eye, regimes with abrupt transitions between them, while 45~present detectable regimes with smoother transitions.

\begin{figure}[t!]
  \begin{center}
  \includegraphics[width=0.98\columnwidth]{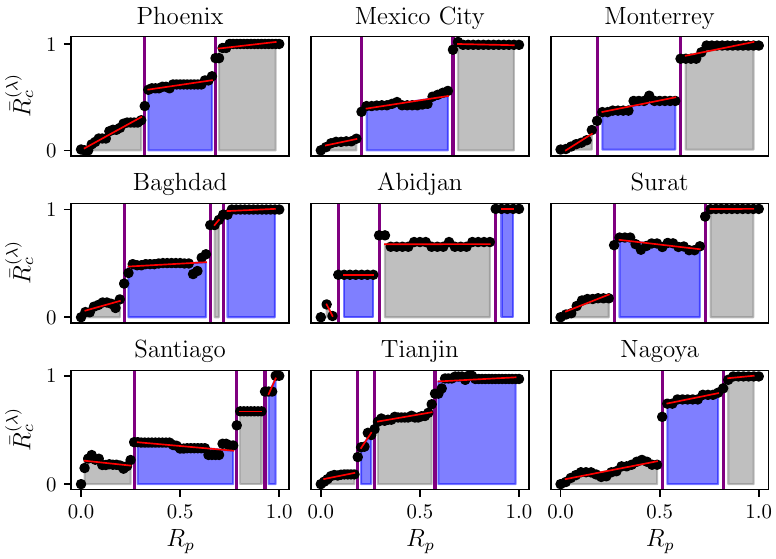}
  \end{center}
  \caption{Analysis of congestion regimes for nine cities. Points represent the average distance from city center of the $\lambda = 15$ nodes with largest betweenness. Vertical lines indicate the change points. Plots are normalized between~0 and~1 considering the radius of the different cities. Regimes and transitions have been automatically detected by the unsupervised method described in Appendix~\ref{sec:elbow}.}
  \label{fig:citiesGTmodel}
\end{figure}

\begin{figure*}[t!]
  \begin{center}
  \includegraphics[width=0.98\textwidth]{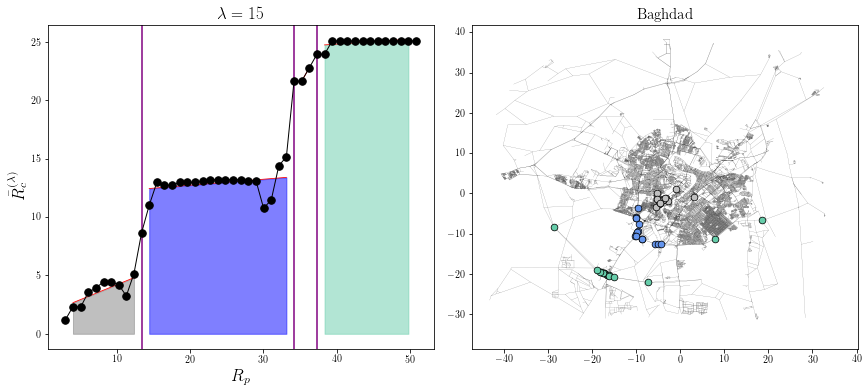}
  \end{center}
  \caption{Spatial behavior of the congestion nodes for the city of Baghdad. In the left we present the dependence of the quantity introduced in Eq.~(\ref{eq:RLambda}) as a function of the patch radius $R_p$, as well as in Fig.~\ref{fig:citiesGTmodel}. It is possible to detect the existence of three congestion regimes, marked with different colors, associated to different city areas where congestion occur. These can be recognized in the right, where we show the map of the city, with the congestion nodes related to each regime. Precisely, for each value of $R_p$ we plot the corresponding $\lambda$ high betweenness nodes, with a color of the regime they belong. In the grey regime, congestion nodes are located in the city center, while in the blue one they falls on its perimeter. For the green one, congestion are mostly located at the connections between the city center and peripheral arterial roads.}
  \label{fig:Baghdad}
\end{figure*}

Figure~\ref{fig:citiesGTmodel} shows the results for 9~of the networks with clear multiple abrupt transitions, with radius ranging from the $39$~km of Abidjan to the $64$~km of Tianjin. Equivalent results for another 21~cities are presented in Fig.~S11 of the Supplemental Material \cite{SM}. For these 30~cities, the behavior is very close to our theoretical predictions. For illustrative purposes and to compare the real city structure with our synthetic models, in Fig.~\ref{fig:Baghdad}, we present the map of the city of Baghdad, highlighting the $\lambda$ highest betweenness nodes associated to each patch. It is possible to recognize the groups of nodes for each congestion regime, either localized in the city center, its perimeter, or in the connections of the city center to the periphery. The similarities with the regimes unveiled by the GT-model are outstanding.

\begin{figure}[tb!]
  \begin{center}
  \includegraphics[width=0.99\columnwidth]{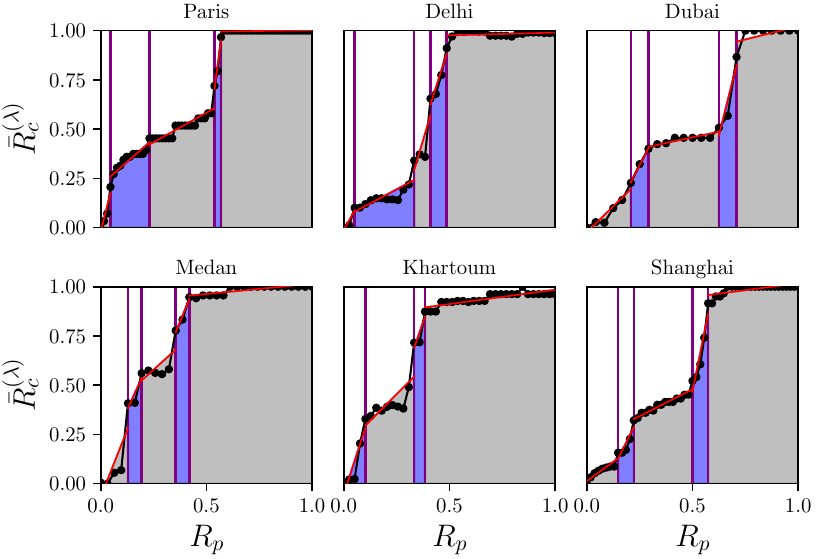}
  \end{center}
  \caption{Several examples of real cities that present softer regime transitions between the identified regimes. Points represent the average distance from city center of the $\lambda = 15$ nodes with largest betweenness. Vertical lines indicate the change points. Plots are normalized between~0 and~1 considering the radius of the different cities. Regimes and transitions have been automatically detected by the unsupervised method described in Appendix~\ref{sec:elbow}.}
  \label{fig:citiesSoftTransitions}
\end{figure}

Regarding the number of congestion regimes detected, our automatic method has shown that the analyzed cities have between~3 and 5~regimes. However, one has to consider that our method looks for changes in mean and slope, thus, in some situations in which the transition is not so abrupt, the transition itself may be identified as a regime. When this happens, the amount of congestion regimes lays between~2 and~3. See Fig.~\ref{fig:citiesSoftTransitions} for several examples of that situation. The amount of detected abrupt transitions is in clear agreement with the models developed in Sec.~\ref{sec:GTmodel} and \ref{sec:RandomPlanarModels}, where either two or three regimes are detected. Anyway, we must remark that our main objective is not to accurately predict the number of transitions, but to point out their existence, and to relate them to the structural discontinuity in urban road networks.

In this regard, it is important to understand the mechanisms that make transitions abrupt. Comparing the left panel of Fig.~\ref{fig:PhaseDiagram} with Figs.~S4 and~S5 of the Supplemental Material \cite{SM}, adding structural noise to our synthetic models turns into shifts and uncertainty of the transitions. Similarly, in the context of the DT-MST model, smoother transitions may also be recovered by relaxing the structural discontinuity between arterial and central roads, i.e., making this structural transition occur in a wider spatial range.

Finally, in Fig.~S12 of the Supplemental Material \cite{SM} we present a sensitivity analysis for different values of $\lambda \in \{2,5,10,15\}$, which shows that the detected pattern persists as well for lower values of $\lambda$. Additionally, in Figs.~S13 and~S8 of the Supplemental Material \cite{SM} we test $\bar{R}^{(\lambda)}_c$ on the GT and DT-MST models, respectively, showing that the abrupt transition patterns persist.

\section{Discussion and Perspectives}
\label{sec:Discussion}

The management and control of congestion has a direct impact on the efficiency and the engineering of urban transportation networks. In this paper, the analysis of a rich dataset composed of 97~road networks of cities worldwide have shown the existence of several congestion regimes, with abrupt transitions, associated to the geographic areas where the principal road bottlenecks occur. Using our GT and DT-MST models, we have shown (analytically and experimentally) that the regimes arise because of the entanglement between the arterial and central roads and their embedding in separated spaces. A transition between city center and periphery appears due to this spatial change in structure, while the existence of a second transition depends on the number of separated arboreal structures that form the periphery. Additionally, the abrupt transitions between the regimes respond to how fast is the structural discontinuity between them, with softer transitions emerging by tightening their inter-connectivity (e.g., in the form of structural noise).

Most of our results are based on the analysis of the location of bottlenecks as we take, incrementally, larger concentric urban areas. Although there is not a direct mapping, we understand that this process may mimic, at least, two different process occurring in urban settings: urban growth, and changes in hourly mobility patterns.

From the structural point of view, as cities evolve, arterial roads connecting different cities are engulfed by the urban sprawl \cite{Strano2012}, which presents structural regularities \cite{han2020classification,marshall2004streets} similar to central parts of our models (grid and DT). This growing process can be simulated with our models by fixing the size of the trees and analyzing its behavior with respect to increasing sizes of the grid. Alternatively, several cities with strong historical background have evolved from a dense, grid-like center, towards a sparse, tree-like structure \cite{masucci2013, Kostof1999} which could also be modeled with our models. Meanwhile from the dynamical point of view, faster transportation means (and, particularly, cars) have contributed to longer commuting distances in geographically spread-out cities, which means that (effectively) cities have extended to wider areas. This is, as well, clearly related to the analysis we provide here, when the city center (with regular structure) is fixed and only the periphery part is extended.

At another time scale, we find daily urban mobility patterns. During morning rush hour, citizens depart from the dormitory cities towards the urban center. In that situation, the effective road network structure spans far from the city center (large patch radius in our analysis). As the day progresses, traffic from dormitory cities is reduced and internal traffic determines the effective transportation network (low value of patch radius). Intermediate patch radius, consider the transition between these two opposite situations.

As seen with the previous examples, although our results may seem theoretical in some aspects, several urban processes can be analyzed within our framework by either fixing the urban center (grid or DT) or the periphery structure (regular tree or MST) and varying the size of the remaining structure.

Taking a wider perspective on the presented results, the general conclusion of the paper can be shortly summarized by stating that critical nodes (bottlenecks) on road networks present a general tendency to be abruptly shifted away from the city center, as cities increasingly incorporate larger areas and conurbations. Although each city may have its own particularities, from a complex networks point of view, we can consider this phenomenon as unexpected but desirable in real urban road networks.

Unexpected, since it is difficult to conclude whether road networks have been designed with this purpose in mind, or it is a byproduct of some other process. In general, one could have also expected that congestion radius increases according to a continuous law, without any abrupt jump.

To see why it is a desirable phenomenon, let us first analyze the dramatic effects if no transitions would exist. We have seen through our analysis that this happens in lattices, regular trees, Delaunay Triangulation, Maximum Spanning Trees (see Fig.~S9 of the Supplemental Material \cite{SM})), and in networks with total overlap between these simple structures (see Fig.~S10 of the Supplemental Material \cite{SM})); that is, in situations where there is no structural differentiation between the center and the periphery of the network. In these arrangements, the location of critical nodes does not depend on the size of the network, meaning that if road networks grow maintaining the same structural properties, then critical nodes (located at the city center) will receive increasing pressure, extremely deteriorating the system. We can then conclude that, as cities grow (or incorporate new peripheral traffic in time), there is a need for their road network to incorporate structural changes in it. Once these are incorporated (as the ones described in this paper), we see that expelling bottlenecks out from the urban center may, in part, avoid severe congestion problems.

Within this context, excessively abrupt displacement of bottlenecks (our detected abrupt structural transitions) indicate points or regions where the structural transition is excessively sharp, and this must have implications, yet to be studied, with respect to the efficiency (in whatever terms) of the transportation system. Although we still have uncertainties about the effect of transition abruptness, our developments have shown that smoother structural transitions can be achieved by tightening the connectivity between the different road types, for instance, adding noise between the two structures (see Fig.~S4 compared to Figs.~\ref{fig:PhaseDiagram} and~S5.).

\section{Conclusions}
\label{sec:Conclusions}

In this work, we have unveiled and studied the existence of multiple abrupt transitions in the location of urban road networks bottlenecks. With our models, we have provided understanding about the essential structural features of road networks that promote the emergence of this phenomenology, and we have given indications on how to control the abruptness of the transitions. Finally, we elaborate on the implications of our work to urban planning.

Despite the extensive analysis we have provided, there is, clearly, much work to do ahead. The incorporation of road traffic data, the analysis of the specific effects of the phase transitions on the routing dynamics, and the development of smart methods to alleviate the abrupt transition, are some exercises that may become necessary to fully understand the effects of the phenomenon described here.

Overall, we believe that our work puts some light on the importance of the entanglement between the different road types, and provides clues on the fundamental reasons of the observed phenomena. Although further research is necessary, the discovered issue is important, and needs to be addressed to optimize the efficiency of urban transport systems.

\section*{Acknowledgements}
We thank A.\ Kirkley, G.\ Ghoshal, H.\ Barbosa and M.\ Barthelemy for sharing the road network data of the cities we analyze. A.L., J.B-H.\ and A.S-R.\ acknowledge the support of the Spanish MICINN project (grant PGC2018-096999-A-I00). A.L.\ acknowledges the support of a postdoctoral grant from the Universitat Oberta de Catalunya (UOC). S.G.\ acknowledges financial support from Spanish MINECO (grant PGC2018-094754-B-C21), Generalitat de Catalunya (grant No.\ 2017SGR-896), and Universitat Rovira i Virgili (grant No.\ 2019PFR-URV-B2-41).

\appendix

\section{GT-model with noise}
\label{sec:noise}

We have tested three types of noise: (1) random edge addition with length bias; (2) random edge removal with length bias; and (3) Delaunay triangulation noise. For all noise models, edges are gradually inserted or removed until graph density $\rho=E/(3N-6)$ is modified in the desired amount $\Delta \rho=\Delta E /(3N-6)$, where $E$ is the number of edges and $N$ the number of nodes of the network. Note that for planar graphs $0\leqslant\rho\leqslant 1$, see Chapter~2 of \cite{barthelemy2018morphogenesis}. Since the noise is introduced with biases proportional to the distances between the nodes, it is necessary to assign first coordinates to the nodes.

\subsection{Planar embedding}
\label{sec:PlanEmbed}
The coordinates origin, i.e., the node with position $(0,0)$, is assumed to be located in the center of the grid, which is well defined since the grid is supposed to have sides with an odd number of nodes, $w=2\ell+1$ (perfect grid), as shown in Fig.~\ref{fig:GTModel}. This holds even for grids with a general size $N_{G}$ (such as in Fig.~\ref{fig:PhaseDiagram}), because they are obtained by randomly adding new nodes to the periphery of a perfect grid. In the grid, nodes coordinates $(x_g,y_g)$ remain defined as the number of jumps required to reach the central node: $x_g$ counts the jumps in the horizontal direction, while $y_g$ in the vertical one.

\begin{figure}[tb!]
  \begin{center}
  \includegraphics[width=0.98\columnwidth]{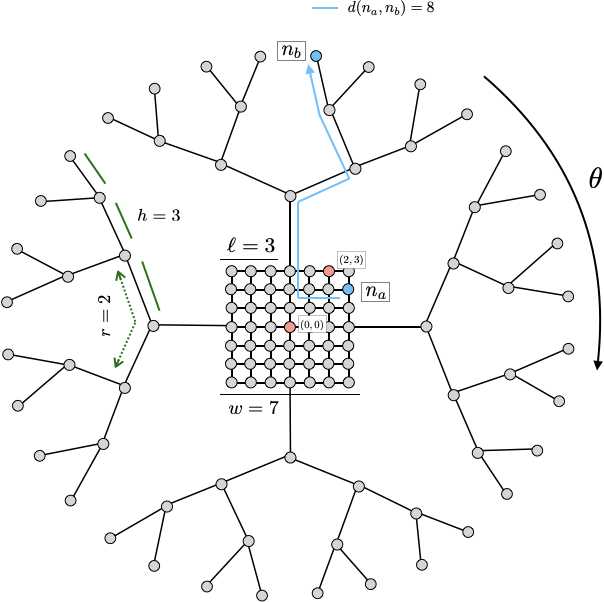}
  \end{center}
  \caption{Example of the planar embedding of the GT-model with parameters $w=7$, $r=2$ and $h=3$. }
  \label{fig:planar}
\end{figure}

It is possible to proceed in a similar way for the tree nodes. Here position of a node $t$ is defined using polar coordinates with respect to the grid-center node:
\begin{equation}
  (x_t, y_t) = (R^{(t)} + 4d_t) (\cos(\theta_t), \sin(\theta_t))\,,
\end{equation}
where
\begin{itemize}
    \item The tree-root is set to be in the same axis as the corresponding connector node, and at a distance $R^{(t)}=\sqrt{2}\,\ell+2$ from the grid center. This selection is useful to avoid the collision between tree and grid nodes;
    \item $d_t$ is the number of jumps required to reach the node $t$ from its tree-root (i.e., the tree level);
    \item $\theta_t$ accounts for the angular separation of node~$t$ with respect to the grid center. All nodes at the same level of the tree, and for the four trees of the GT-model, are uniformly distributed along the circle of radius $R^{(t)} + 4d_t$.
\end{itemize}
An example of this planar embedding is shown in Fig.~\ref{fig:planar}.

\subsection{GT-model with additive length bias noise}
\label{sec:noiseadditive}

For this type of noise, we randomly add edges with a probability inversely proportional to a certain power of the euclidean distance $d_{ij}$ between endpoints~$i$ and~$j$. The iterative procedure works as follows: for a given pair of nodes~$i$ and~$j$ chosen uniformly at random, an edge $(i,j)$ is added with probability $1-(d_{ij})^{-\epsilon}$, provided that planarity restrictions are not compromised. The value of~$\epsilon>0$ controls the bias towards the introduction of new edges.

\subsection{GT-model with negative length bias noise}

This kind of noise is implemented following the same procedure as for additive noise, with the only difference that now edges are removed, rather than added.

\subsection{GT-model with Delaunay noise}

This type of noise is generated considering the Delaunay triangulation of the GT-model nodes, mimicking the procedure in \cite{Kirkley2018}. Once the triangulation is generated, edges, uniformly chosen at random, are gradually inserted to the base model until the desired edge density is reached.

\section{Betweenness of connector nodes of the GT-model}
\label{sec:Connector}

The calculation of the betweenness of the connector nodes of the GT-model, i.e., the nodes in the center of the sides of the grid to which the root of the trees are connected, can be done following the same approach as in Sec.~\ref{sec:AnBeetwCalc}. First, we decompose the betweenness in three contributions using Eq.~(\ref{eq:BmaxGTModel}):
\begin{equation}
  B^{(c)} = a^{(c)} N^2_{T} + b^{(c)} N_{T} + c^{(c)}\,.
  \label{eq:Bconnector}
\end{equation}
Let us consider for example the connector node to tree~$T_4$ in Fig.~\ref{fig:GTModel}. All the shortest paths coming from nodes in the other three trees have to pass through it, thus
\begin{equation}
  a^{(c)} = 3\,.
\end{equation}

Similarly, all the paths starting in nodes of the grid and going to the tree~$T_4$ also cross our connector node, contributing to its betweenness with one unity per origin node, i.e., with a term~$w^2-1$. Additionally, this connector also participates in other paths between the grid and the other trees. More precisely, if the origin is a node in the same side of the grid as the connector, there are two trees with shortest paths through the connector. For example, if we take as origin the node which is at a distance~$y$ below the considered connector in Fig.~\ref{fig:GTModel}, there is one path crossing the connector among the $\pi_{2\ell,y}$ shortest paths to arrive to tree~$T_2$, and $\pi_{\ell,\ell}$ paths through the connector among the $\pi_{\ell,\ell+y}$ to arrive to tree~$T_1$. Given the up-down symmetry of the system, the final expression for coefficient $b^{(c)}$ is
\begin{equation}
  b^{(c)} = w^2-1 + 2\sum_{y=1}^{\ell} \left( \frac{1}{\pi_{2\ell,y}} + \frac{\pi_{\ell,\ell}}{\pi_{\ell,\ell+y}} \right).
\end{equation}

Finally, shortest paths that start and end in the grid and cross a connector node need again that one of the endpoints is in the same side as the connector, thus we choose again as origin a node at a distance~$y$ below the previously considered connector. The destination node can be identified with variables $(a,b)$, the horizontal (to the left) and vertical (upwards) distances to the connector node, respectively. The case $a=0$ consists of destinations in the same side as the origin and connector, thus contributing to the betweenness of the connector with a value $\ell^2$. When $a>0$ there are $\pi_{a,b}$ paths through the connector node among the total $\pi_{a,b+y}$ shortest paths connecting the origin and destination nodes, thus we get:
\begin{equation}
  c^{(c)} = \ell^2 + 2 \sum_{a=1}^{2\ell} \sum_{b=0}^{\ell} \sum_{y=1}^{\ell}  \frac{\pi_{a,b}}{\pi_{a,b+y}}\,.
  \label{eq:c_c}
\end{equation}

\section{Betweenness of tree nodes of the GT-model}
\label{sec:FullCompleteTree}

The symmetries of full and complete trees allow for the calculation of the betweenness of all their nodes. In fact, all nodes of the tree located at the same level share the same value of the betweenness, thus we can denote it as~$B(v)$, where level~$v\in\{0,\ldots,h\}$ is~$0$ for the root and~$h$ for the leaves of the tree.

Trees are characterized by the absence of cycles. As a consequence, there is a unique path connecting every pair of nodes, which means that all $\sigma_{od}=1$, thus simplifying Eq.~(\ref{eq:SPBdef}).

Let us consider a node at level~$v$ of the tree. There are two types of paths that cross it: paths that come from one children (level~$v+1$) and continue to one of its siblings (level~$v+1$); paths that come from the parent (level~$v-1$) and continue to one of the children (level~$v+1$). We need to count how many different paths exist for each of these types.

In the first case, for each of the $r(r-1)/2$ pairs of children, there are $N_{T}(h-v-1)$ possible origins of the path, and the same number of possible destinations, thus forming a total of $\frac{r(r-1)}{2} N_{T}(h-v-1)^2$ different paths. Here, we have made use of Eq.~(\ref{eq:sizes}), which provides the number of nodes in a full and complete tree with branching ratio~$r$ and height~$h$:
\begin{equation}
  N_{T}(h) = \frac{r^{h+1}-1}{r-1}\,.
\end{equation}
In our case, we have applied it to calculate the number of nodes of the sub-tree formed by a child from level~$v+1$ and all its descendants.

In a similar way, the number of paths that cross the node and its parent is equal to the number of descendants of the node, $r\,N_{T}(h-v-1)$, multiplied by the number of the rest of the nodes, $N_{T}(h)-N_{T}(h-v)$. Combining both results, the betweenness of a node at level~$v$ reads:
\begin{align}
  B(v) =\ &\frac{r(r-1)}{2} N_{ T}(h-v-1)^2 \nonumber
  \\
  &+ \left[N_{T}(h)-N_{T}(h-v)\right]r\,N_{T}(h-v-1)\,,
  \label{eq:Bv1}
\end{align}
which can be written as:
\begin{equation}
  B(v) = \frac{r(r^{h-v}-1)}{2(r-1)^2}
  \left(2 r^{h+1}-r^{h-v+1}-r^{h-v}-r+1\right).
\end{equation}
This expression is valid for all nodes of the tree, including the root ($v=0$). In particular, betweenness vanishes for the leaves, $B(h)=0$, as expected.

It is worth noting that maximum betweenness of the tree is located at the root only for trees with branching ratio $r>2$; binary trees have the maximum at the children of the root. This can be shown by calculating the difference of betweenness between levels $v=0$ and $v=1$:
\begin{equation}
  B(0)-B(1) = \frac{r^h}{2(r-1)}
  \left[r^{h-1}(r^2-2r-1)+2\right].
  \label{eq:B0mB1}
\end{equation}
The term in brackets is negative if $r=2$ and $h>2$. To obtain Eq.~(\ref{eq:B0mB1}), we have made use of the property
\begin{equation}
  N_{T}(h)=N_{T}(h-1)+r^h\,.
\end{equation}

Once we have determined the betweenness within the tree, we need to include the contribution of the rest of the network which forms the GT-model. The new paths to consider are those starting outside the tree, with destination in a node descendant of the one for which we want to calculate the betweenness. If this node belongs to level~$v$, following the same procedure which has led to Eq.~(\ref{eq:Bv1}), the result is
\begin{equation}
  B_{total}(v) = B(v) + \left[N_{ GT}-N_{T}(h)\right]r\,N_{T}(h-v-1)\,.
  \label{eq:Btotal}
\end{equation}
Note that $N_{GT}=w^2+4N_{T}(h)$. Now, we obtain the final expression for the desired betweenness of the tree-roots of the GT-model, $B^{(t)}=B_{total}(0)$:
\begin{equation}
  B^{(t)} = \frac{r(r^{h}-1)}{2(r-1)^2}
  \left[ 7r^{h+1}-r^h+(2w^2-1)(r-1)-6\right].
  \label{eq:Btreeroot1}
\end{equation}

For binary trees ($r=2$), the consideration of the full GT-model network makes the root of the tree become the node of maximum betweenness among the rest of the nodes in the tree:
\begin{align}
  B_{ total}&(0)-B_{total}(1) =
  \nonumber
  \\
  &\frac{r^{h}}{2(r-1)}
  \left[ r^{h-1}(7r^2-2r-1)+2w^2(r-1)-4 \right].
\end{align}
This time, the term in brackets is positive for all values of the branching ratio, height of the tree, and size of the grid.

The betweenness of the tree-root nodes can also be expressed in terms of the sizes of the trees of the GT-model, as in Eq.~(\ref{eq:BmaxGTModel}), if we do not expand the value of $N_{T} = N_{T}(h)$ in Eqs.~(\ref{eq:Bv1}) and~(\ref{eq:Btotal}):
\begin{equation}
  B^{(t)} = \frac{r(r-1)}{2} \left( \frac{N_{ T}-1}{r} \right)^2 + (w^2 + 3N_{T}) (N_{ T}-1)\,,
  \label{eq:Btreeroot2}
\end{equation}
where we have made use of
\begin{equation}
  N_{T}(h-1) = \frac{N_{T} - 1}{r}\,.
\end{equation}
Rearranging the terms in Eq.~(\ref{eq:Btreeroot2}) we get
\begin{equation}
  B^{(t)} = a^{(t)} N^2_{T} + b_j^{(t)} N_{T} + c_j^{(t)}\,,
  \label{eq:Btreeroot}
\end{equation}
with
\begin{align}
  a^{(t)} &= 3 + \frac{r-1}{2r}\,,
  \\
  b^{(t)} &= w^2 - 3 - \frac{r-1}{r}\,,
  \\
  c^{(t)} &= \frac{r-1}{2r} - w^2\,.
\end{align}

\section{Congestion regimes for the GT-model with non-complete grids}
\label{sec:noncompletegrid}

Analytical betweenness expressions in Sec.~\ref{sec:AnBeetwCalc} have been derived assuming the case of complete grids, i.e., grids with side $w=2\ell+1$ ($n\in\mathbb{N}$) and $w^2$ nodes. In Fig.~\ref{fig:PhaseDiagram}, we also introduced incomplete grids, to fill the gaps between grids of odd-squared sizes. These incomplete grids, with extra nodes in the periphery, modify differently the values of the betweenness of the connector and tree-root nodes, thus shifting the transition between these regimes.

\begin{figure}[tb!]
  \begin{center}
  \includegraphics[width=0.98\columnwidth]{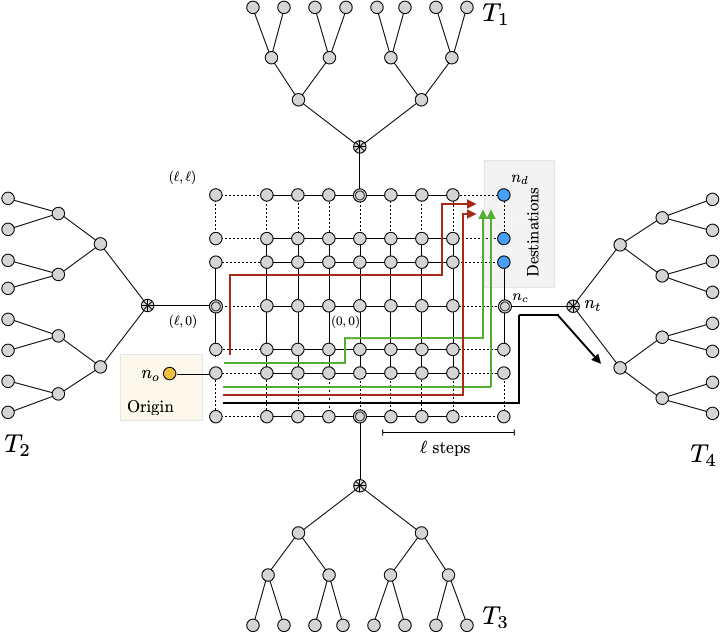}
  \end{center}
  \caption{Graphical representation of a network generated with the GT-model with parameters $w=2\ell + 1$, $r=2$ and $h=3$, in which an additional node $n_o$ has been attached to periphery. Colors, labels and notation are set to explain the derivation in Appendix~\ref{sec:noncompletegrid}.}
  \label{fig:noncompletegrid}
\end{figure}

Consider Fig.~\ref{fig:noncompletegrid} as an illustration of the following process. We start by connecting a new node $n_o$ to the left side of a complete grid, and quantify its contribution to the betweenness of the connector node $n_c$ (belonging to the side in front of the new node), and the adjacent tree-root, $n_t$. After adding $n_o$, nodes $n_c$ and $n_t$, among others, experience an increment in its betweenness that can be formalized as:
\begin{eqnarray}
  \Delta B^{(t)} & = & N_{T} - 1\,,
  \\
  \Delta B^{(c)} & = & \Delta B^{(t)} + 1 + \delta^{(c)}\,,
\end{eqnarray}
The value of $\Delta B^{(t)}$ includes the contribution to betweenness of paths between $n_o$ and the nodes of the tree to which node $n_t$ belongs to. Since all paths need to cross $n_t$ to reach their destinations, this value is equivalent to the number of nodes in the tree minus one. These paths also cross the connector and contribute to $\Delta B^{(c)}$, see the black path in Fig.~\ref{fig:noncompletegrid}. However, $\Delta B^{(c)}$ embodies a new term, $\delta^{(c)}$, that considers the paths between the new node and the grid-ones located in the same side as the connector $n_c$, and above it, see the green paths in Fig.~\ref{fig:noncompletegrid}. It turns out that
\begin{equation}
  \Delta B^{(c)} > \Delta B^{(t)},
\end{equation}
which explains why, in Fig.~\ref{fig:PhaseDiagram}, numerical simulations associated to the connector regime overcome the solid black frontier defined by Eq.~(\ref{eq:NumTAn}). Formally, we may write
\begin{equation}
  \delta^{(c)}=\frac{\sigma_{n_o,n_d}(n_c)}{\sigma_{n_o,n_d}}\,.
\end{equation}
It can be approximated by
\begin{equation}\label{eq:CorrPaths}
  \delta^{(c)} \lesssim 2\sum_{b=1}^{\ell} \sum_{y=1}^{\ell} \frac{\pi_{w,b}}{\pi_{w,b+y}}\,,
\end{equation}
where we have taken advantage of the same idea used to derive the $c^{(c)}$ term in Eq.~(\ref{eq:c_c}). This is an approximation because we are supposing that the side in which node $n_o$ is located is also full of new nodes, above and below it. This gives a good upper bound for most configurations, which is enough to obtain the corrected transition boundary depicted as a dashed line in the right panel of Fig.~5.

\section{Detection of the number of congestion regimes}
\label{sec:elbow}

\begin{figure}[tp!]
  \begin{center}
  \includegraphics[width=0.98\columnwidth]{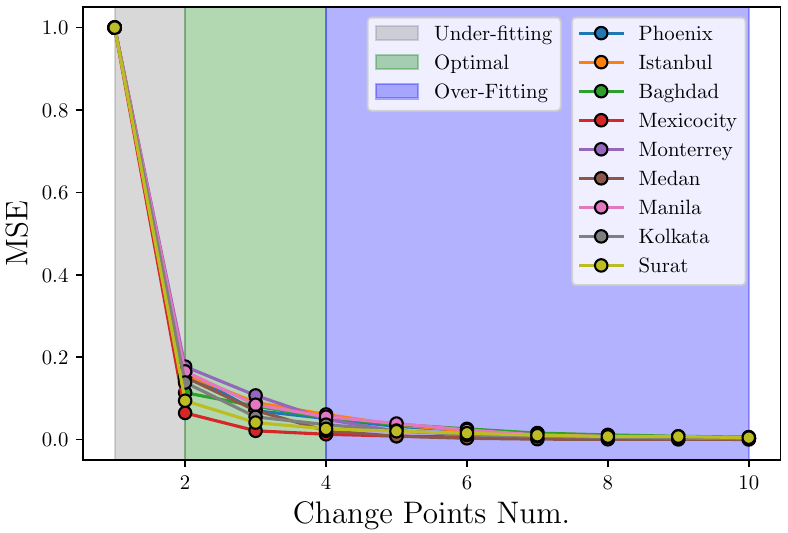}
  \end{center}
  \caption{Relationship between fitting error (MSE) and the number of change points used to approximate the curves in Fig.~\ref{fig:citiesGTmodel}}
  \label{fig:elbow}
\end{figure}

Theoretical analysis based on GT-model predicts three different phases that can be clearly defined by two cut points (a.k.a.\ change points). To identify these pattern in empirical road networks (see Fig.~\ref{fig:citiesGTmodel}), one has to decide the number of change points to consider. Note that, as the number of change points increases, the fitting error decreases. To automatically decide the optimal number of change points, we recall to the elbow test introduced in \cite{ElbowRef}. The method is based on the concept of diminishing returns to balance the accuracy obtained with respect to the number of change points considered.

In Fig.~\ref{fig:elbow} we plot the evolution of the mean squared error (MSE) of the fit with respect to the number of change points. As expected, the resulting line draws an elbow: in general, the error rapidly decreases as the number of change points grows, but after a certain value, adding another change point doesn't provide much improvement. Such a surplus of precision can be considered a kind of over-fitting. Of course, the opposite situation, namely a very low number of change points, leads to under-fitting. Thus, the elbow singles out the optimal number of change points. In Fig.~\ref{fig:elbow} we see that the elbow position approximately falls between~2 and~4 change points, endorsing the GT-model prediction.


%

\onecolumngrid
\newpage

\newcounter{SMfigure}
\renewcommand{\thefigure}{S\arabic{SMfigure}}

\stepcounter{SMfigure}
\begin{figure}[ht!]
  \begin{center}
  \includegraphics[width=0.98\columnwidth]{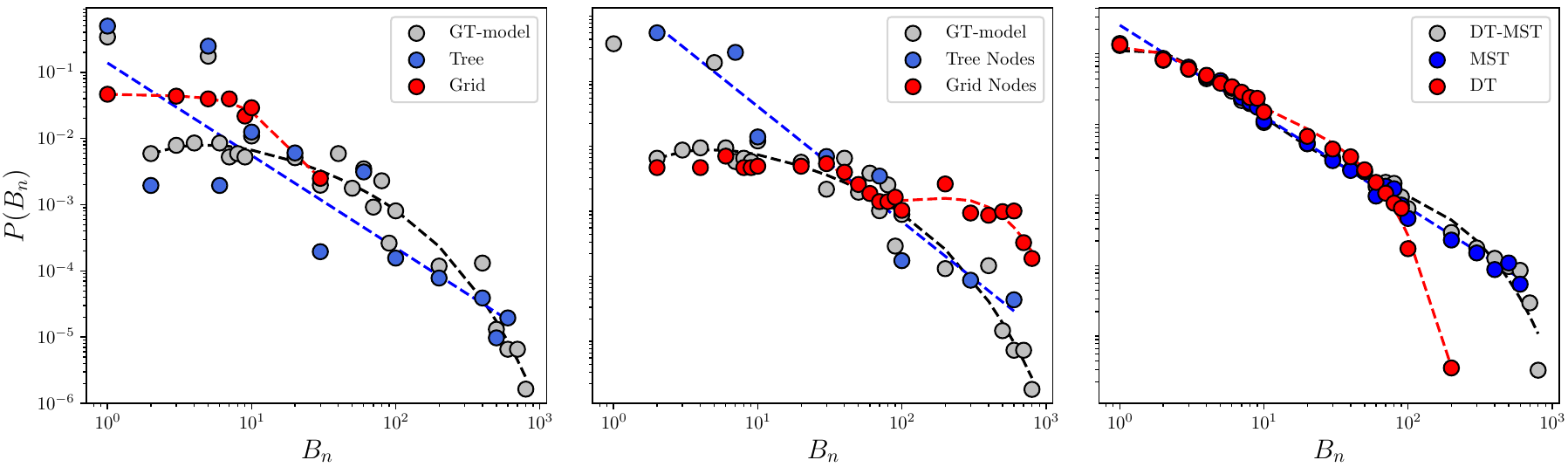}
  \end{center}
  \caption{Betweenness distribution comparison of the GT-model, DT-MST, and the related null models: grid, tree, DT and MST. Left: A GT-model with $w=45$, $h=10$, and $r=2$ is compared with its individual components, i.e., a grid with $w=45$, and a tree with $h=10$, $r=2$.   Middle: The same GT-model is compared with the betweenness distribution resulting from its grid and tree nodes. Right: A DT-MST model (obtained from the GT-model above, according the procedure explained in Sec.~VI, and with $\Delta N/N=1$) is compared with its underlying DT and MST structures. In all the cases, the betweenness values are normalized by the number of nodes. Dashed lines represent the resulting polynomial fit and highlight the cut-off role of the grid/DT and the damping one of the trees. In the left and right plots, the null models refer to independent graphs, while in the center we consider the grid and tree node contributions to the same GT-model network.}
  \label{fig:NullModels_Panel}
\end{figure}

\stepcounter{SMfigure}
\begin{figure}[ht!]
  \begin{center}
  \includegraphics[width=0.90\columnwidth]{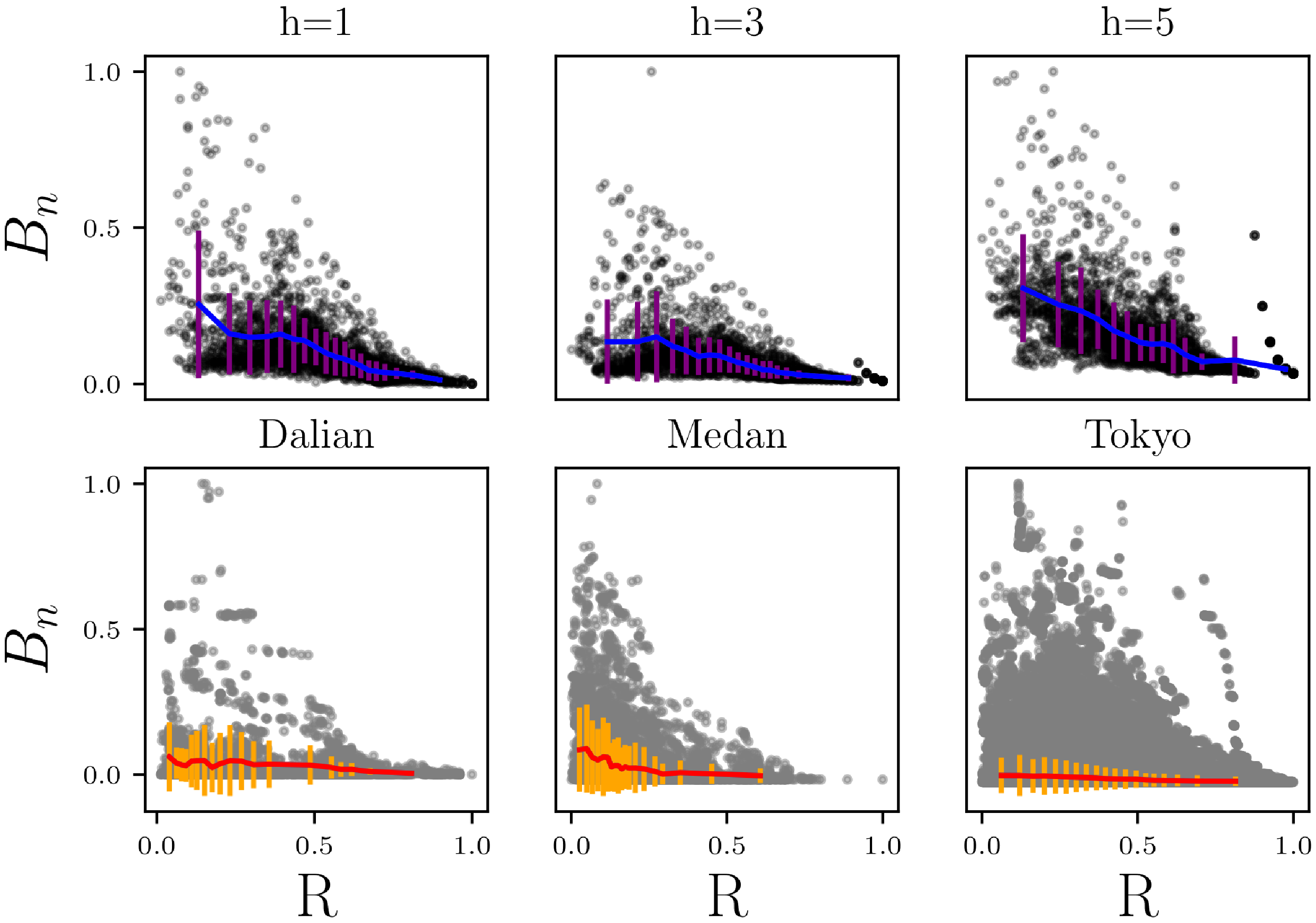}
  \end{center}
  \caption{Comparison between the real (bottom) and estimated (top) relationship between nodes geographic position and their betweenness for small (Dalian, China), medium (Medan, Indonesia) and big cities (Tokyo, Japan). The GT-models used have parameters $w=51$, $r=2$, $\Delta\rho/\rho=1.5\%$, and the values of $h$ indicated in each column. All distances and betweenness are normalized between 0 and 1. Coordinates for the nodes of the GT-model have been assigned using the planar embedding described in Appendix~A1. Solid lines on the scatter plots represent the mean betweenness at the given radius $R$. Deviations correspond to one $\sigma$ of the distribution.}
  \label{fig:GTmodelSPBdist}
\end{figure}

\stepcounter{SMfigure}
\begin{figure}[ht!]
  \begin{center}
  \includegraphics[width=0.98\columnwidth]{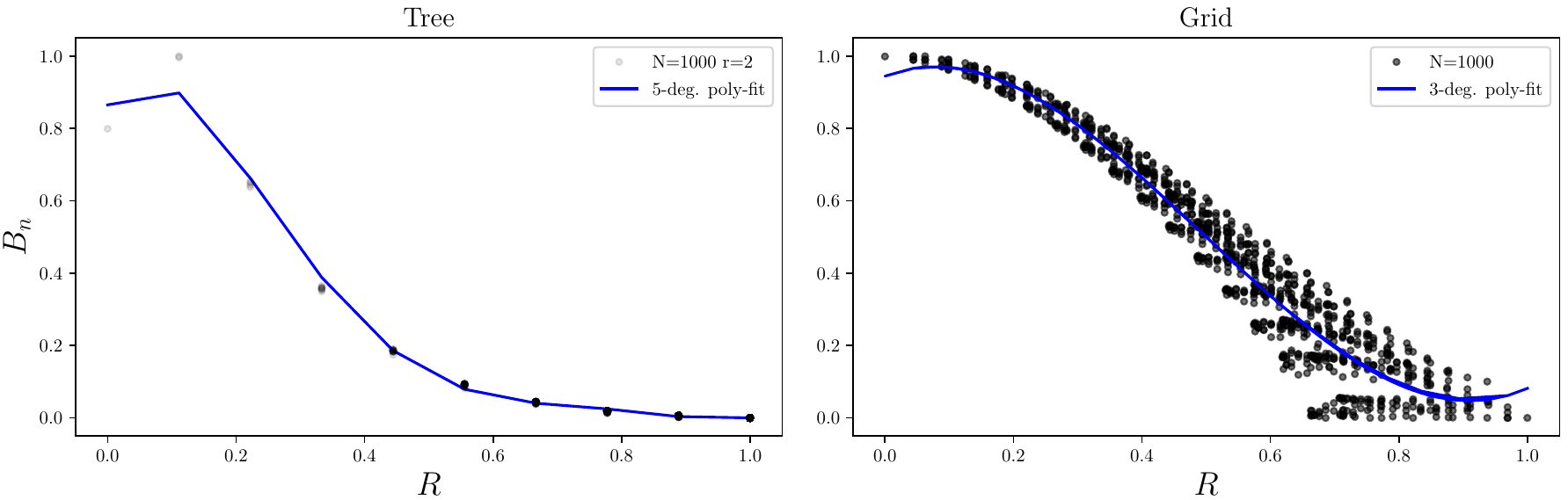}
  \end{center}
  \caption{Spatial behavior of the betweenness centrality for a tree (left) and a grid (right). Each point represents a node, and its horizontal and vertical coordinates correspond to its betweenness and geographic position, respectively. In the case of the grid, geographic position is defined as the euclidean distance from its geometric center, while it refers to the number of jumps from its root in the case of the tree. The blue line portrays a polynomial fit of the betweenness radius behavior. \vspace{3cm}}
  \label{fig:NullModels_SpatialBehavior}
\end{figure}

\stepcounter{SMfigure}
\begin{figure}[ht!]
  \begin{center}
  \includegraphics[width=0.85\columnwidth]{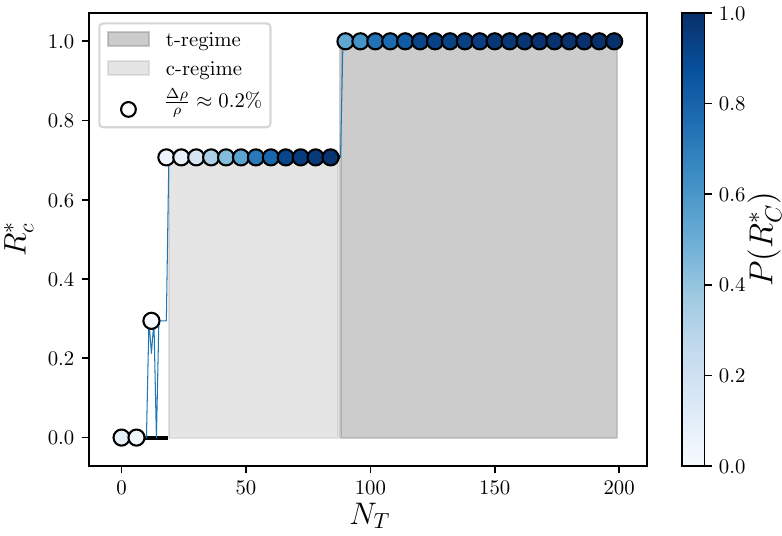}
  \end{center}
  \caption{Influence of noise in the prediction of the average congestion radius $R_c$ as a function of $N_{T}$, the number of nodes of the trees, with fixed branching factor $r=2$ and grid width $w=25$, and a noise level of $0.2\%$. Values are averaged over an ensemble of $n=150$ realizations of the GT-model. Solid black lines present regimes as predicted by Eq.~(17). Circles present the experimental results after the addition of noise. Each circle is located at the statistical mode obtained with the distribution of $R_c$ after the 150~realizations. The color of the circle shows the probability of that value over the experimentally obtained distribution of $R_c$.}
  \label{fig:noisyTransition}
\end{figure}

\stepcounter{SMfigure}
\begin{figure}[ht!]
  \begin{center}
  \includegraphics[width=0.85\columnwidth]{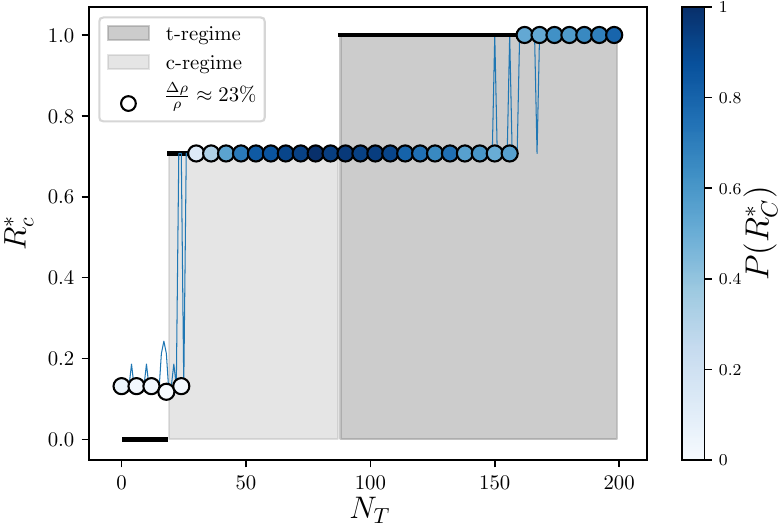}
  \end{center}
  \caption{Influence of noise in the prediction of the average congestion radius $R_c$ as a function of $N_{T}$, the number of nodes of the trees, with fixed branching factor $r=2$ and grid width $w=25$, and a noise level of $23\%$. Values are averaged over an ensemble of $n=150$ realizations of the GT-model. Solid black lines present regimes as predicted by Eq.~(17). Circles present the experimental results after the addition of noise. Each circle is located at the statistical mode obtained with the distribution of $R_c$ after the 150~realizations. The color of the circle shows the probability of that value over the experimentally obtained distribution of $R_c$. }
  \label{fig:noisyTransition2}
\end{figure}

\stepcounter{SMfigure}
\begin{figure}[ht!]
  \begin{center}
  \includegraphics[width=0.85\columnwidth]{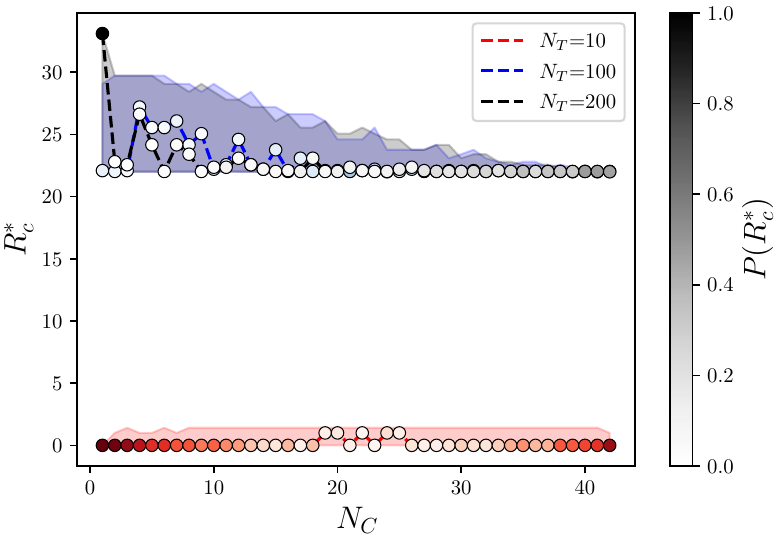}
  \end{center}
  \caption{Dependence of the congestion radius on $N_C$, i.e., the number of peripheral trees connected to each side of the grid. All the trees are assumed to have the same number of nodes, and the point in which they are connected to the grid side is randomly chosen. Each point is located at the statistical mode (here indicated by $R^{*}_c$) resulting from the $R_c$ distribution after $100$ graph realizations at fixed $N_C$, and its color provides the occurrence of the $R^{*}_c$ value in the distribution. Shadow areas fill the space between the minimum and maximum value taken by the congestion radius over the ensemble. Trees are connected to a grid characterized by $w=45$ and exhibit branching factor $r=2$, while their number of nodes is presented in the legend. The $N_T$ are chosen in order to cover the three regimes. It is possible to see that the difference between the value the congestion radius takes in the t and c-regimes vanishes as the number of connections grows. However, it is always possible to distinguish a congestion center regime and a peripheral one.}
  \label{fig:EntryPoints1}
\end{figure}

\stepcounter{SMfigure}
\begin{figure}[ht!]
  \begin{center}
  \includegraphics[width=0.85\columnwidth]{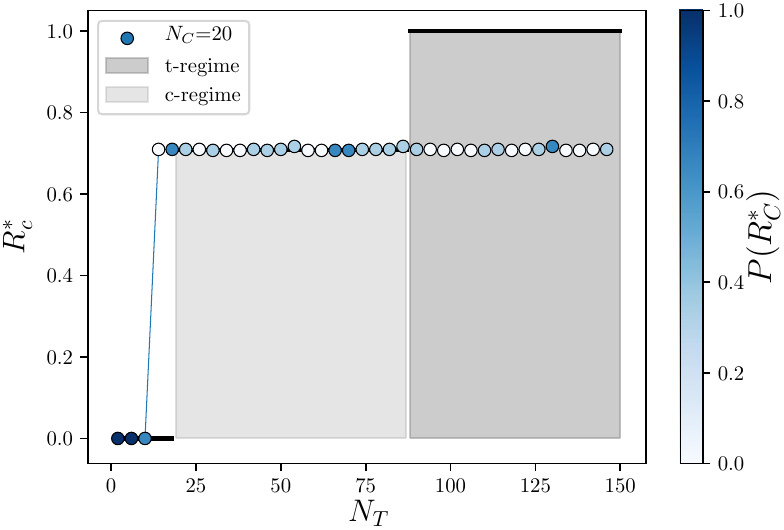}
  \end{center}
  \caption{Dependence of the congestion radius on the number of tree nodes in the case of $N_C=20$ peripheral trees. Trees are connected by the root to a random node of the grid perimeter. Each circle is located at the statistical mode (here indicated by $R^{*}_c$) resulting from the $R_c$ distribution after $20$ graph realizations at fixed $N_C$, and its color provides the occurrence of the $R^{*}_c$ value in the distribution. Trees are connected to a grid characterized by $w=25$ and branching factor $r=2$. The t and c regimes are not distinguishable through a measure of the congestion radius for any value of $N_T$, however the abrupt transition between the g and c regimes still holds.}
  \label{fig:EntryPoints2}
\end{figure}

\stepcounter{SMfigure}
\begin{figure}[ht!]
  \begin{center}
  \includegraphics[width=0.94\columnwidth]{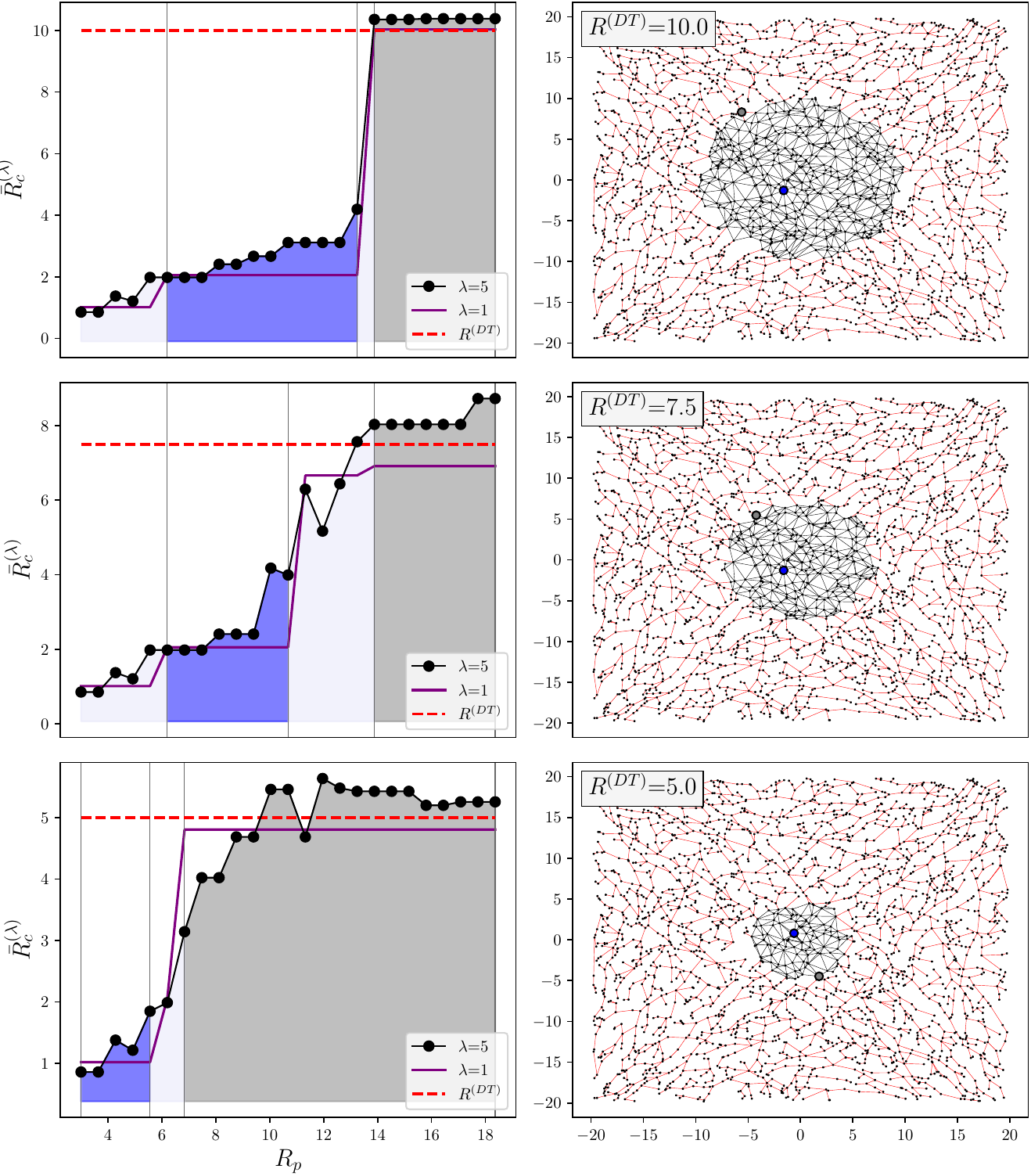}
  \end{center}
  \caption{Spatial behavior of the congestion nodes for different configurations of the random planar model. The location of the nodes is completely random, i.e., $\Delta N/N=1$. The left column shows the dependence of the congestion radius, evaluated by means of the quantity $\bar{R}^{(\lambda)}_c$ introduced in Eq.~(19), as a function of the patch radius. The horizontal red dashed line represents the value of $R^{DT}$, separating the DT area from the MST one. In the right column we depict the corresponding network configuration, where DT and MST edges are painted in red and black, respectively. Grey and blue points are the first and second more frequent congestion nodes emerging for all the patches, and the corresponding patch radius range is painted with the same color in the left column.}
  \label{fig:DT_MST_model_radius}
\end{figure}

\stepcounter{SMfigure}
\begin{figure}[ht!]
  \begin{center}
  \includegraphics[width=0.98\columnwidth]{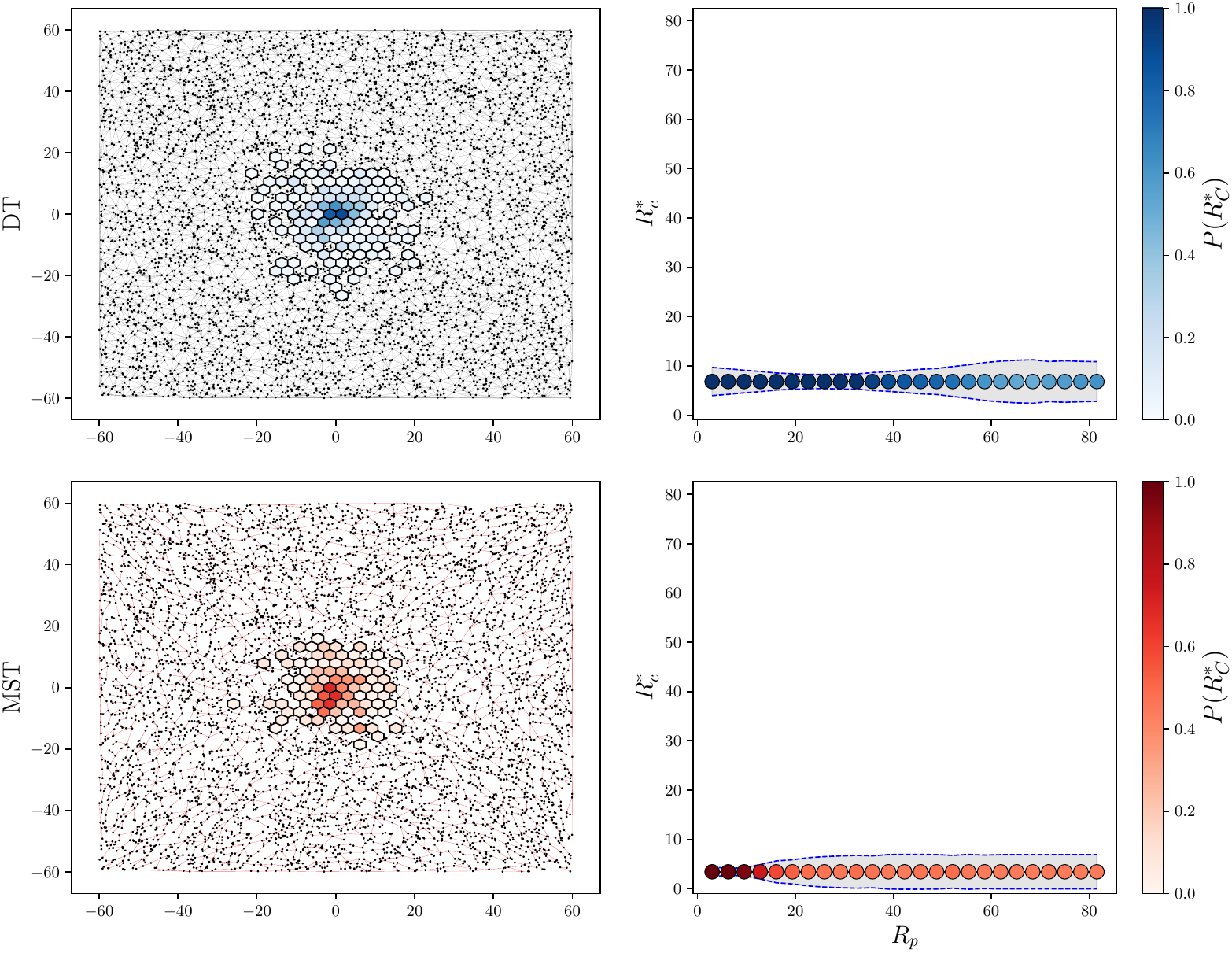}
  \end{center}
  \caption{Spatial behavior of the congestion nodes for the DT (top) and the MST (bottom). The location of the nodes is completely random. The figure is organized similarly to Fig.~6: left column shows the network configuration, where hexagonal bins describe the occurrence of congestion in space, and right column presents the dependence of congestion radius on the patch radius. Here, circles are located at the statistical mode over an ensemble of 100~realizations of different random distributions of $N=6000$ nodes in a square of side equal to $\ell=60$. Importantly, congestion mostly occurs in the center and no transitions arise, suggesting that these are a consequence of the intertwining of different graph structures.}
  \label{fig:NullModels}
\end{figure}

\stepcounter{SMfigure}
\begin{figure}[ht!]
  \begin{center}
  \includegraphics[width=0.98\columnwidth]{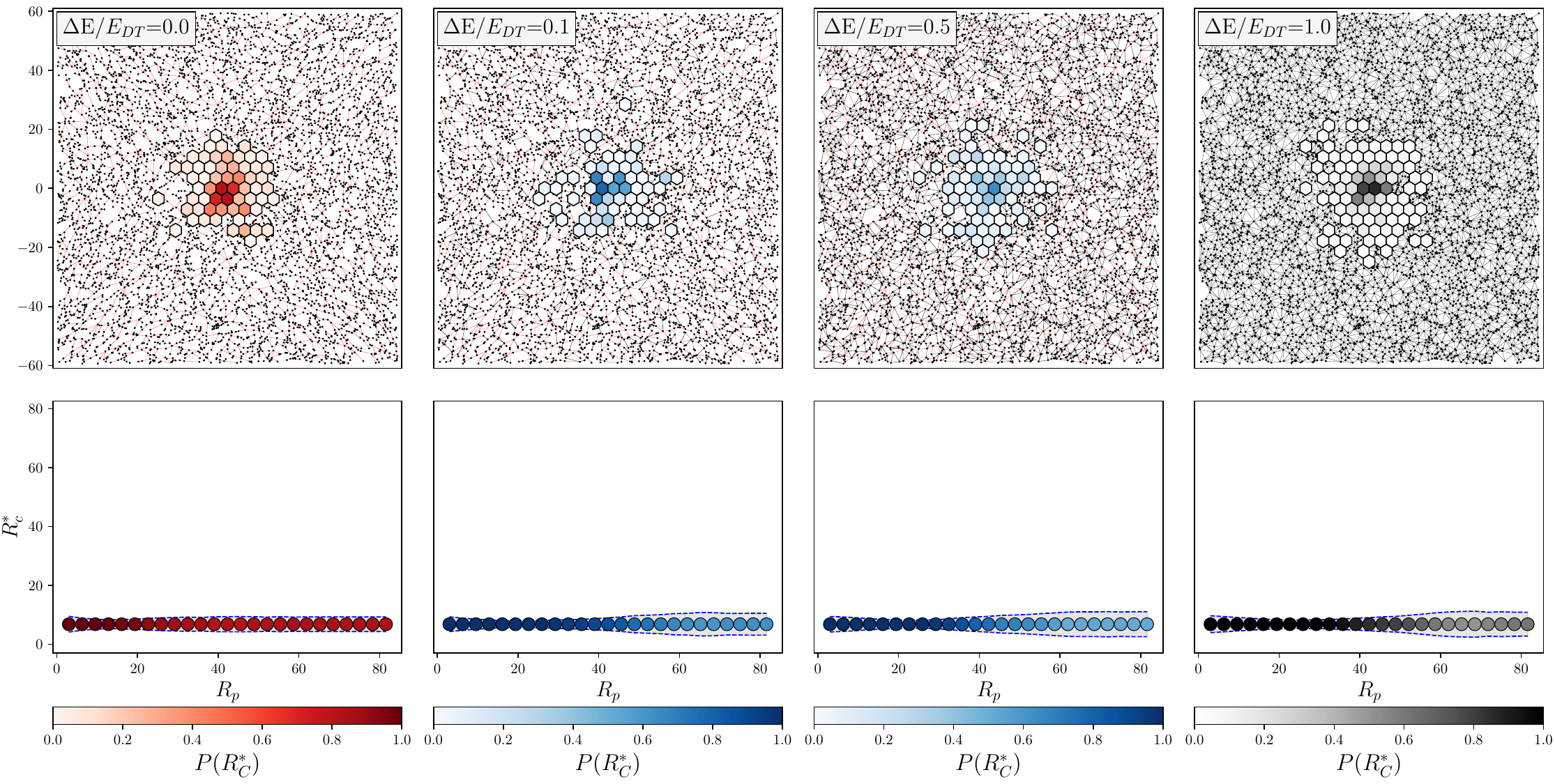}
  \end{center}
  \caption{Spatial behavior of the congestion nodes for random planar models without any spatial separation between DT and MST. A DT is calculated for all the randomly  located nodes, then its MST based on edge betweenness is found, and all the edges of this MST are included in the network. From all the DT edges not in the MST, a randomly selected fraction $\Delta E/E_{DT}$ is also added to the network (Delaunay noise, as explained in Appendix~A). The figure is organized in the same manner as Fig.~6: top row shows the network configuration, where hexagonal bins describe the occurrence of congestion in space, and bottom row presents the dependence of congestion radius on the patch radius. Here, circles are located at the statistical mode over an ensemble of 100 realizations of the edge noise  over the same random node distribution of $N=6000$ nodes in a square of side equal to $\ell=60$. Still, similarly to Fig.~\ref{fig:NullModels}, congestion mostly occurs at the center and no transitions arise, suggesting that the spatial separation between DT and MST is a necessary condition for them.}
  \label{fig:NoSpatialSep}
\end{figure}

\stepcounter{SMfigure}
\begin{figure}[ht!]
  \begin{center}
  \includegraphics[width=0.99\columnwidth]{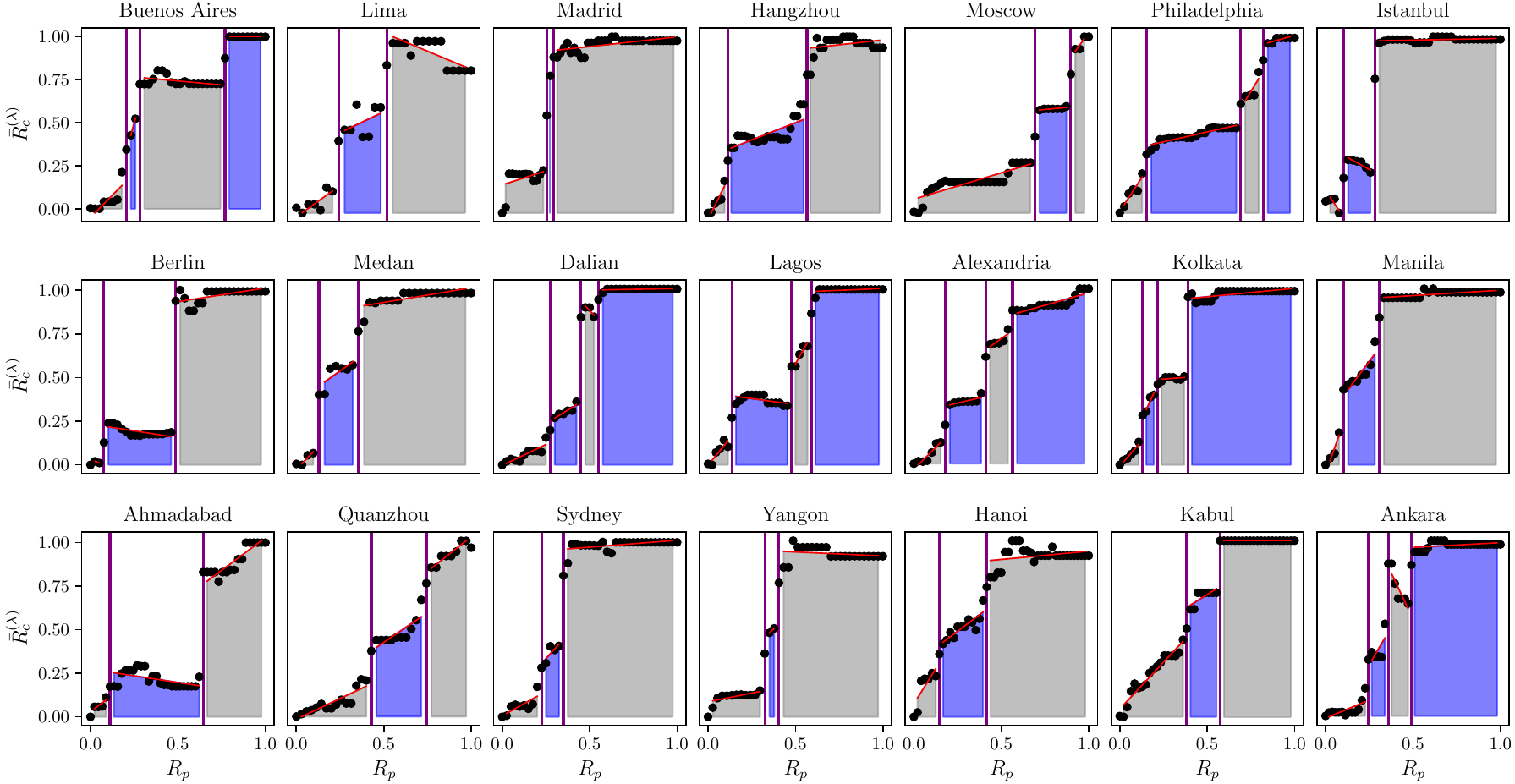}
  \end{center}
  \caption{Additional analysis of congestion regimes on a set of 21~cities. Each point in the different plots represents the average distance from city center of the $\lambda = 15$ maximum betweenness nodes. Vertical lines indicate the change points, i.e., where either the mean or the slope of the congestion radius changes. Plots are normalized between 0 and 1 considering the radius of the different cities.}
  \label{fig:citiesGTmodelSM}
\end{figure}

\stepcounter{SMfigure}
\begin{figure}[ht!]
  \begin{center}
  \includegraphics[width=0.95\columnwidth]{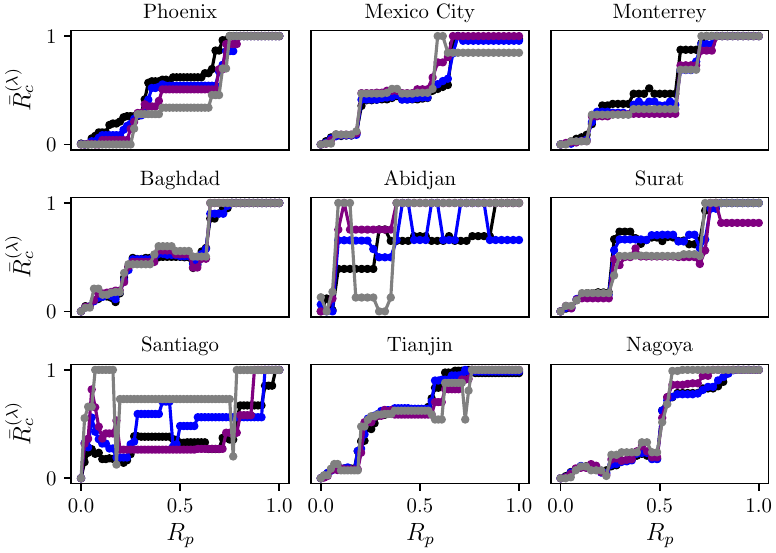}
  \end{center}
  \caption{Analysis of the congestion regimes on several cities depending on $\lambda$. Points represent the average distance from city center of the $\lambda$ maximum betweenness nodes: black line $\lambda=15$; blue line $\lambda=10$; purple line $\lambda=5$; grey line $\lambda=2$.}
  \label{fig:citiesGTmodelLambda}
\end{figure}

\stepcounter{SMfigure}
\begin{figure}[ht!]
  \begin{center}
  \includegraphics[width=0.85\columnwidth]{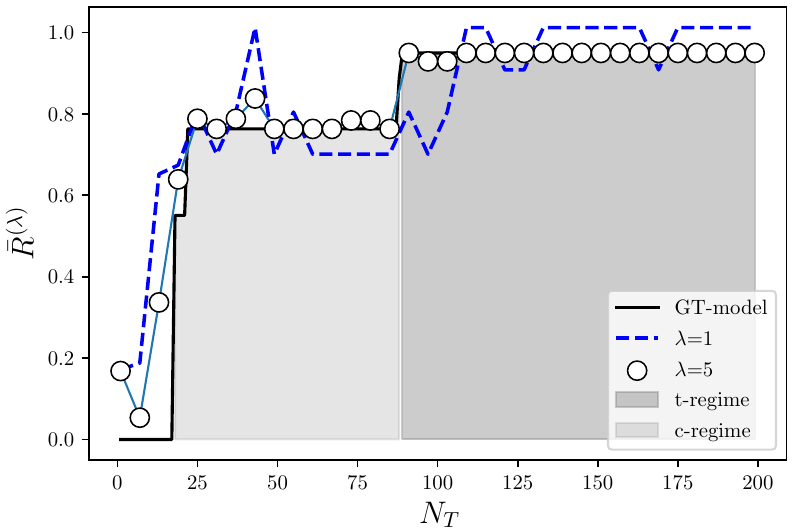}
  \end{center}
  \caption{Validation of the quantity $\bar{R}^{(\lambda)}$ introduced in Eq.~(19) on the GT-model, with fixed values of the branching factor $r=2$ and grid width $w=25$, as a function of the number of nodes of the trees, $N_{T}$. Solid black line refers to the GT-model without noise, while circles present the experimental results after the addition of noise with $\Delta\rho/\rho=0.2\%$. In these two cases, the value of $\lambda$ is set to be equal to 5, while the blue dashed line represents the same noisy behavior with $\lambda=1$. In particular, we consider $n=20$ realizations of the biased edge addition noise. Here, the final value of $\bar{R}^{(\lambda)}$ is averaged over these realizations. It results that the employ of $\bar{R}^{(\lambda)}$ preserves the abrupt transitions pattern detected by means of the $R_c$ distribution mode, as shown for instance in Figs.~\ref{fig:noisyTransition} and~\ref{fig:noisyTransition2}. The difference between the values of $\bar{R}^{(\lambda)}$ in c and t regimes gets smaller as $\lambda$ increases; this may be understood since, for $\lambda\geqslant 8$, the definition in Eq.~(19) covers both the tree roots and middle-side grid nodes.}
  \label{fig:RlambdaGTmodel}
\end{figure}

\end{document}